\documentclass[acmtog, authorversion=True]{acmart}
\acmSubmissionID{0335}

\usepackage{array}
\usepackage{booktabs} 
\usepackage{multirow}  
\usepackage{enumerate}

\usepackage{algorithm}
\usepackage[noend]{algorithmic}

\setcopyright{acmcopyright}\acmJournal{TOG}
\acmYear{2021}\acmVolume{40}\acmNumber{4}\acmArticle{153}\acmMonth{8}
\acmDOI{10.1145/3450626.3459821}

\usepackage{xspace}

\newcommand{\rev}[1]{#1}

\definecolor{pink}{rgb}{1,0,1}
\newcommand{\revisiondelete}[1]{}

\newcommand{\reals}{\ensuremath{\mathbb{R}}}
\newcommand{\integers}{\ensuremath{\mathbb{Z}}}
\newcommand{\integersNonNeg}{\ensuremath{\integers^{+}}}

\newcommand{\denselist}{\itemsep 0pt\parsep=0pt\partopsep 0pt\vspace{-\topsep}}

\newcommand{\methodname}{\text{ShapeMOD}\xspace}

\begin{document}
\title
[\methodname: Macro Operation Discovery for 3D Shape Programs]
{\methodname: Macro Operation Discovery for 3D Shape Programs}

\author{R. Kenny Jones}
\affiliation{%
    \institution{Brown University}
}

\author{David Charatan}
\affiliation{%
    \institution{Brown University}
}

\author{Paul Guerrero}
\affiliation{%
    \institution{Adobe Research}
}

\author{Niloy J. Mitra}
\affiliation{%
    \institution{University College London and Adobe Research}
}

\author{Daniel Ritchie}
\affiliation{%
    \institution{Brown University}
}

\begin{abstract}

A popular way to create detailed yet easily controllable 3D shapes is via procedural modeling, i.e. generating geometry using programs. 
Such programs consist of a series of instructions along with their associated parameter values.
To fully realize the benefits of this representation, a shape program should be compact and only expose degrees of freedom that allow for meaningful manipulation of output geometry.
One way to achieve this goal is to design higher-level \emph{macro} operators that, when executed, expand into a series of commands from the base shape modeling language.
However, manually authoring such macros, much like shape programs themselves, is difficult and largely restricted to domain experts.
In this paper, we present \methodname, an algorithm for automatically discovering macros that are useful across large datasets of 3D shape programs.
\methodname operates on shape programs expressed in an imperative, statement-based language. 
It is designed to discover macros that make programs more compact by minimizing the number of function calls and free parameters required to represent an input shape collection. 
We run \methodname on multiple collections of programs expressed in a domain-specific language for 3D shape structures. 
We show that it automatically discovers a concise set of macros that abstract out common structural and parametric patterns that generalize over large shape collections. 
We also demonstrate that the macros found by \methodname improve performance on downstream tasks including shape generative modeling and inferring programs from point clouds.
Finally, we conduct a user study that indicates that \methodname's discovered macros make interactive shape editing more efficient.

\end{abstract}

%
%

\ccsdesc[500]{Computing methodologies~Computer graphics}
\ccsdesc[500]{Computing methodologies~Neural networks}
\ccsdesc[500]{Computing methodologies~Probabilistic reasoning}

%
%

\keywords{Shape analysis, shape synthesis, generative models, deep learning, procedural modeling, neurosymbolic models}

\begin{teaserfigure}
  \centering
  \includegraphics[width=\linewidth]{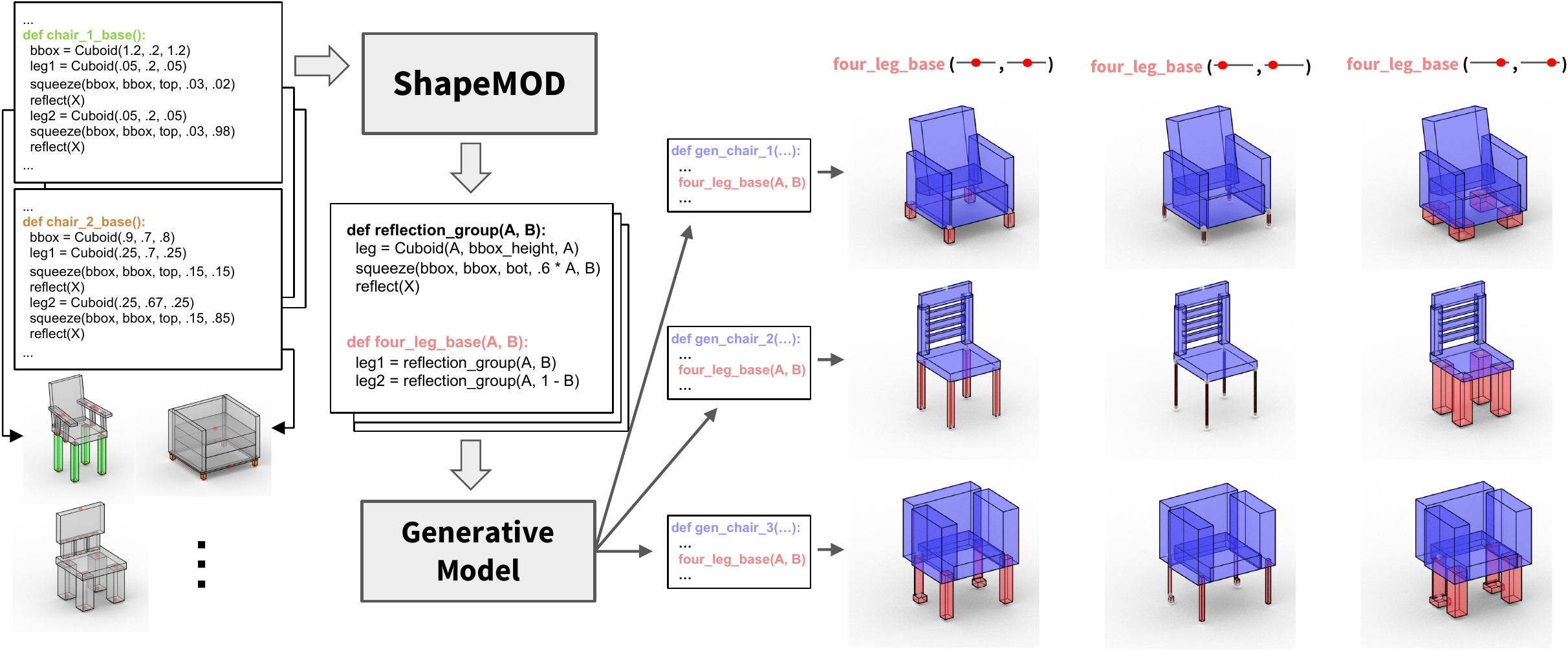}
\caption{
We propose \methodname, an algorithm which takes as input a collection of 3D shape programs and makes them more compact by automatically discovering common \emph{macros} which can be re-used across the collection.
\rev{We apply \methodname to datasets of ShapeAssembly programs} and find that
generative models which train on refactored programs containing these macros produce more plausible output shapes than those trained on the original programs.
The discovered macros also facilitate shape editing by exposing only a small number of meaningful parameters for manipulating shape attributes. For example, the \emph{four\_leg\_base} macro exposes two parameters (visualized as sliders with red handles); one parameter controls leg size, while the other controls leg spacing. 
}
\label{fig:teaser}
\end{teaserfigure}

\maketitle

\section{Introduction}
\label{sec:intro}

3D data is important for many applications in graphics, vision, and artificial intelligence: creating content for games and AR/VR experiences, synthetic training data for autonomous agents, etc.
Unlike 2D image data, which has a standard representation as a regular grid of pixels, 3D data can be expressed in a variety of representations, each with their own strengths and weaknesses.
Point clouds are easy to obtain from the real world via depth sensors, but they lose surface connectivity and detail; 
meshes are well-supported by rendering and simulation packages, but they are harder to train machine learning models on; 
implicit functions provide excellent surface detail but are harder to inspect and reason about.

One interesting way to represent a 3D shape is with a program that generates its geometry, or at least its high-level structure.
This is particularly appealing for manufactured objects, as such shapes typically \emph{originate} as CAD programs of some form (e.g., connected assemblies of parts, the geometry of which may be specified by lower-level instructions).
In computer graphics, the procedural modeling literature offers a rich history demonstrating the value of programs for shape modeling.
By using simple domain-specific languages (DSLs) that combine geometric elements, programs (such as shape grammars) can model shapes from a variety of domains, such as plants~\cite{LSystemsBook}, buildings~\cite{CGAShape}, cities~\cite{CityEngine}, and furniture~\cite{FabricableFurnitureGrammar}.
In addition to enabling rapid creation of a variety of content, programs offer the benefit of being more readable and editable (via the free parameters they expose) than other shape representations.

To maximally realize these benefits, a good shape program should be compact and expressed at a high level while still exposing important degrees of freedom for editing.
One way to create such programs is to introduce higher-level functions, or \emph{macros}, into the shape DSL.
We define a macro to be a function that, when executed, expands into a series of commands from the base DSL.

In urban procedural modeling, a macro might capture how primitives combine to make a particular class of railing; in furniture modeling, a macro might be used to model a shelving pattern that could be instantiated within different types and sizes of furniture; in plant modeling, a macro might be used to instantiate examples of petal structures across a family of flowers.
In past work on modeling 3D shape structures, rudimentary macros proved to be helpful for downstream tasks of interest, such as shape editing and generative modeling~\cite{jones2020shapeAssembly}. 
However, these macros were carefully (manually) designed by experts and may not generalize to or be useful for other shape modeling domains for which they were not designed.
One question naturally follows: \textit{Can useful macros for a given shape modeling domain be designed automatically by an algorithm?}

In this paper, we present \methodname, an algorithm for automatically discovering such macros from a collection of shape programs.
\methodname~operates on any imperative, statement-based language whose commands are parameterized by discrete and continuous parameters.
It is designed around the principle of discovering macros that make programs more \emph{compact}, where compactness is measured by the number of function calls and number of free parameters required to represent the input shape collection.

In pursuit of compactness, one must consider the cost incurred by adding more functions (i.e., macros) to the DSL.
At one extreme, one could use no macros, which results in the maximum number of free parameters (i.e., minimal compactness).
At the other extreme, one could define a macro for each shape program in the input collection---this is maximally compact, but makes applications such as shape manipulation or learning to generate novel shape programs impossible.
Our insight is that the trade-off space between these extremes can be navigated via optimization to find a middle-ground where a small set of macros explain a high percentage of variations across the input collection of shape programs.
Critically, these frequently-used macros expose sufficient degrees of freedom to allow for shape manipulation and exploration across a shape collection.

We run \methodname~on multiple collections of shape programs expressed in the ShapeAssembly DSL~\cite{jones2020shapeAssembly} to discover new libraries of macros. 
For example, in Figure~\ref{fig:teaser}, starting from a set of chair shape programs, \methodname discovers a reusable macro for four leg chair bases which exposes a compact set of associated control sliders.
We demonstrate the benefits of working with these discovered macros, by evaluating how adding the discovered macros into the language affects performance on downstream tasks: learning a generative model for shape programs, learning to infer shape programs from unstructured geometry, and goal-directed editing of shapes via their programs.
In all cases, task performance is improved by using automatically discovered macros.
\rev{Finally, we show that \methodname~can find useful macros even when trained on a set of ShapeAssembly programs from multiple categories. }

In summary, our contributions are:
\begin{enumerate}[(i)]
    \denselist
    \item An algorithm that takes as input a collection of programs expressed in a DSL with imperative functions that may contain continuous parameters, and automatically discovers a concise set of macros that abstract out common structural and parametric patterns within the input programs that generalize over a shape collection. 
    \item Demonstrations on collections of manufactured objects that using discovered macros leads to better performance on important downstream tasks such as novel shape generation, directed shape manipulation, and inferring shape structures from unstructured geometry (i.e., point clouds).  
\end{enumerate}

\rev{Code can be found at https://github.com/rkjones4/ShapeMOD . Our implementation is not specific to ShapeAssembly, making it possible to apply \methodname to collections of programs in other imperative languages with real-valued parameters. }

\section{Background \& Related Work}
\label{sec:relatedWork}

Our work is related to prior work in program synthesis, automatic program abstraction, shape abstraction, and 3D shape generative modeling.
We also build heavily upon the ShapeAssembly shape structure modeling DSL~\cite{jones2020shapeAssembly}, so we briefly describe its salient features in this section as well.

\subsection{Related Prior Work}

\paragraph{Finding common abstractions over a set of programs}
Our macro-discovery goal is one instance of a more general class of problems: finding common abstractions within a set of programs.
Context-free grammars have been a common class of programs used in prior work on procedural modeling, and grammar induction is one example of common abstraction finding.
Prior work has designed induction algorithms for building facade grammars~\cite{FacadeInduction}, more general grammars~\cite{BayesianGrammarInduction}, and even grammar-like functional programs~\cite{BayesianProgramMerging}.
One more recent system even supports programs with continuous parameters~\cite{ProcmodLearn}.
However, these approaches are all limited to context-free languages.

Recently, another line of work has investigated common abstraction discovery for general functional programs: the Exploration-Compression (EC) algorithm~\cite{ExplorationCompression} and its successor, DreamCoder~\cite{ellis2020dreamcoder}.
Both operate by repeatedly invoking two algorithm phases.
In EC, the two phases are ``exploration'' (trying to find programs that solve input problems) and ``compression'' (finding abstractions common to these programs).
In DreamCoder, these two phases are respectively called ``wake'' and ``sleep.''
Our method, \methodname, also uses a two-phase approach: ``proposal'' (finding candidate abstractions) and ``integration'' (choosing candidate abstractions to add to the language).
EC and DreamCoder operate on functional programs, and while they do provide some support for real-valued free parameters, they are unable to discover parametric relationships between continuous variables. 

\rev{To the best of our knowledge, \methodname is the first approach that discovers macro operations capturing parametric relationships between continuous parameters across a collection of imperative programs. While specialized data structures such as version spaces and E-graphs can efficiently reason about rewrites of functional programs, they cannot efficiently reason over semantic line re-orderings of imperative programs (i.e. maintaining correct execution behavior) and thus are not applicable to languages such as ShapeAssembly.}

\paragraph{Rewriting a single program}

\methodname~discovers macros which are common to a collection of programs.
A related, but different, problem is that of finding modifications that improve a \emph{single} program of interest.
There has been prior work on this problem in the realm of shape modeling languages.
The Szalinski system takes a low-level CAD program as input and searches for a more compact, higher-level program which produces the same output geometry~\cite{Szalinksi}.
The Carpentry Compiler is a program optimizer that finds rewrites of low-level instructions to maintain high-level semantics while optimizing to reduce manufacturing cost \cite{CarpentryCompiler}. 
Our approach can also be seen as related to systems for more general program-rewriting, such as optimizing compilers~\cite{EqualitySaturation}.

\paragraph{Shape abstraction}
\methodname discovers macros by \emph{abstracting out} common patterns in shape programs.
Prior work has investigated how to abstract shapes themselves, either by simplifying the geometry of individual shapes~\cite{AbstractionOfManMadeShapes}, finding common abstractions for collections of shapes~\cite{ShapeCoAbstraction}, or learning to represent shapes via \rev{a union of simple primitives}~\cite{tulsiani2017learning,AdaptiveHierarchicalCuboidAbstraction, deng2020cvxnet, genova2019learning, zou20173d}.
These methods attempt to abstract the \emph{geometry} of shapes, whereas as we are interested in abstracting the procedural structures of programs that generate shape geometry.

\paragraph{Generative Models of 3D Shapes}
One of the major downstream applications we target is learning generative models of 3D shapes.
Many such models have been proposed.
One category of models learns to generate entire objects as unstructured geometry, representing shapes with occupancy grids~\cite{3DGAN}, surface patches~\cite{AtlasNet}, point clouds~\cite{PointCloudGAN}, or implicit functions~\cite{IMNet}.
In contrast, \emph{structure-aware} models learn to generate objects as arrangements of their component parts~\cite{GRASS,StructureNet,gao2019sdmnet,yang2020dsmnet,jones2020shapeAssembly,EG2020STAR,sigg_structureAware:13}.
In our results, we show that training a structure-aware generative model on ShapeAssembly programs, refactored with  discovered macros, improves novel shape generation.

\paragraph{Visual Program Induction}
Another downstream application we target is visual program induction: inferring a program representation of an object given only unstructured sensor input (e.g., an image or a point cloud).
Prior work in this area has targeted custom DSLs~\cite{ellis2018learning,tian2019learning}, constructive solid geometry (CSG)~\cite{InverseCSG,sharma2018csgnet,ellis2019write,LEST} or other CAD representations~\cite{Fusion360Gallery}.
We show that when inferring ShapeAssembly programs from point clouds, targeting a macro-enhanced version of the language leads to better reconstruction quality.

\subsection{Background: ShapeAssembly}

To demonstrate \rev{the benefits of}~\methodname, in this paper we apply it to collections of ShapeAssembly programs \cite{jones2020shapeAssembly}. ShapeAssembly is a domain-specific language that creates 3D shape structures by defining parts and attaching them together. It is executed imperatively, and its functions use both continuous and discrete parameters. To discover more useful macros, we modify the grammar of ShapeAssembly slightly, as described in Appendix \ref{sec:mod_sa_grammar}. 

In its original form, ShapeAssembly contains two base functions and three macro functions. The base functions are \texttt{Cuboid}, which creates an cuboid part proxy, and \texttt{attach}, which moves a Cuboid to satisfy the described spatial relationship. The \texttt{squeeze}, \texttt{reflect} and \texttt{translate} commands are expert-defined macros that abstract common patterns of structural and geometric variation.
Each of these macros expands into a sequence of lower-level \texttt{Cuboid} and \texttt{attach} commands. ShapeAssembly contains both discrete parameters (cuboid IDs, cuboid faces, symmetry axes, etc.) and continuous parameters (cuboid dimensions, attachment points, etc.). Please refer to the original ShapeAssembly paper~\cite{jones2020shapeAssembly} for further details of the language specification.

\section{Macro Operator Discovery}
\label{sec:method}

\begin{figure}[t!]
  \includegraphics[width=0.85\linewidth]{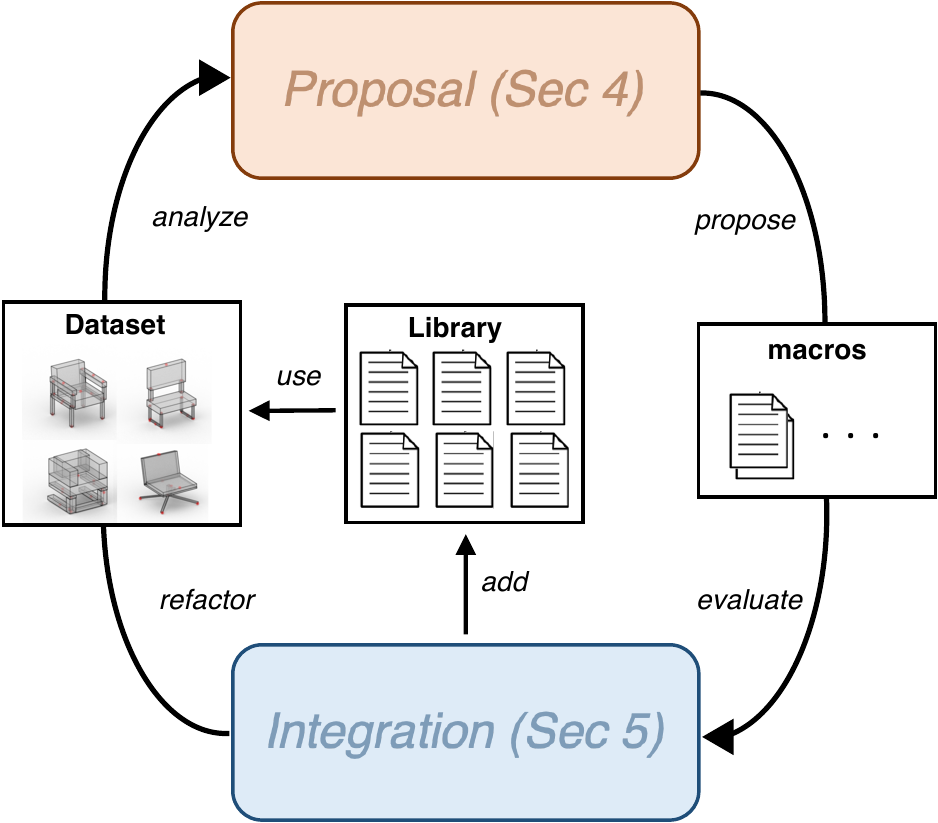}
\caption{
\methodname~consists of two alternating phases: proposing new candidate macros (top) and refactoring programs to use some of the proposed macros (bottom).
}
\label{fig:overview}
\end{figure}

\begin{figure*}[t!]
  \includegraphics[width=\linewidth]{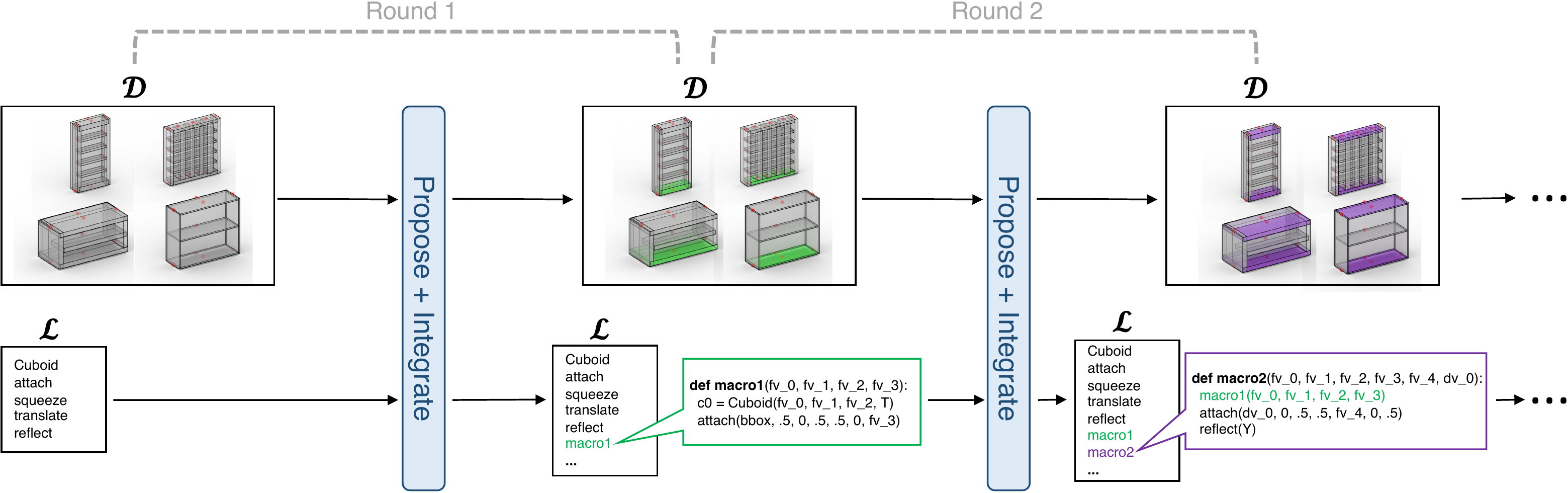}
\caption{
Running \methodname~for multiple rounds allows for discovery of increasingly complex macros.
Here, a macro discovered in Round 2 uses a macro previously found in Round 1 as part of its function body.
}
\label{fig:macro_method}
\end{figure*}

\newcommand{\library}{\ensuremath{\mathcal{L}}}
\newcommand{\dataset}{\ensuremath{\mathcal{D}}}
\newcommand{\objective}{\ensuremath{f}}
\newcommand{\program}{\ensuremath{P}}
\newcommand{\programs}{\ensuremath{\mathcal{P}}}
\newcommand{\order}{\ensuremath{o}}
\newcommand{\orders}{\ensuremath{\mathcal{O}}}
\newcommand{\macro}{\ensuremath{M}}
\newcommand{\macros}{\ensuremath{\mathcal{M}}}

\newcommand{\Tau}{\mathrm{T}}

\definecolor{proposalColor}{rgb}{0.77,0.35,0.066}
\definecolor{integrationColor}{rgb}{0.18,0.33,0.59}

\newcommand{\pluseq}{\mathrel{+}=}

\begin{algorithm}[t!]
\begin{flushleft}
\textbf{Input:} Library of functions $\library$, Program dataset $\dataset$, Objective $\objective$ \\
\textbf{Output:} Updated $\library$ with macros, best programs $\programs^*(\dataset, \library)$\\
\begin{algorithmic}[1]
 \FOR{\emph{num\_rounds}}
 \color{proposalColor}
 \STATE \emph{candidate\_macros} $\gets$ Set()  \hfill \COMMENT{Proposal Phase}
 \FOR{\emph{num\_proposal\_steps}}
 \STATE $\program$, $\order \gets$ sampleProgAndOrder($\dataset$) \hfill \COMMENT{Sec~\ref{sec:form_cluster}}
 \STATE $\programs_\textbf{matches} \gets$ findMatchingProgs($\dataset$, $\program$, $\order$)
 \STATE $\programs_\textbf{cluster} \gets$ sampleByParamSim($\programs_\textbf{matches}$)
 \STATE $\program_\textbf{abs} \gets$ findAbstractProg($\programs_\textbf{cluster}, \library$) \hfill \COMMENT{Sec~\ref{sec:abs_prog}}
 \STATE $\macros \gets$ proposeMacrosForProg($\program_\textbf{abs}$) \hfill \COMMENT{Sec~\ref{sec:cand_funcs}}
 \STATE $\macros \gets$ generalize($\macros$) \hfill \COMMENT{Sec~\ref{sec:gen_prog}}
 \STATE \emph{candidate\_macros} $\pluseq \macros$
 \ENDFOR
 \color{integrationColor}
 \STATE $\tilde{\dataset} \gets$ subsample($\dataset$) \hfill \COMMENT{Integration Phase}
 \FOR {\emph{num\_integration\_steps}}  
 \STATE $\macro \gets$ getTopRankedMacro(\emph{candidate\_macros}) \hfill \COMMENT{Sec~\ref{sec:rank_cand_abs}}
 \STATE $\library' \gets$ optimize($\objective$, $\library$, $\library + \{\macro\}$, $\tilde{\dataset}$) \hfill \COMMENT{Sec~\ref{sec:eval_cand_abs}}
 \STATE $\library \gets \library'$; \textbf{continue}
 \STATE $\macros_\text{infreq} \gets$ findInfrequentMacros($\tilde{\dataset}$, $\library$, $\library + \{\macro\}$)
 \STATE \textbf{continue}
 \STATE $\library' \gets$ optimize($\objective$, $\library$, $\library + \{\macro\} - \macros_\text{infreq}$, $\tilde{\dataset}$)
 \STATE $\library \gets \library'$
 \FOR {$\macro \in \macros_\text{infreq}$}
 \STATE $\library \gets$ optimize($\objective$, $\library$, $\library + \{\macro\}$, $\tilde{\dataset}$)
 \ENDFOR
 \ENDFOR
\FOR {$\macro \in \library$}
 \STATE $\library \gets$ optimize($\objective$, $\library$, $\library - \{\macro\}$, $\tilde{\dataset}$)
\ENDFOR
\STATE $\dataset \gets$ filterBadOrders($\objective$, $\dataset$, $\library$) \hfill \COMMENT{Sec~\ref{sec:filter_bad_orders}}
\ENDFOR
\color{black}
\STATE return $\library$, $\programs^*(\dataset, \library)$ \hfill \COMMENT{Sec~\ref{sec:best_prog}}

\end{algorithmic}
\end{flushleft}

\caption{\methodname}
\label{algo:mod}

\end{algorithm}

\methodname's goal is to take a dataset of programs $\dataset$ and the library of DSL functions used to express them $\library$, and return a new library (with additional macros) which is able to express the programs in $\dataset$ with fewer free parameters.
The motivation here is that macros should remove free parameters that correspond to \emph{extraneous} degrees of freedom, i.e. degrees of freedom that can create implausible output shapes, such as independently changing the length of each leg of a table.
At the same time, we want to keep the number of functions in our library relatively small, so as not to remove \emph{necessary} degrees of freedom that can create meaningful shape manipulations. 
We formalize this trade-off in an objective function $\objective$ which the algorithm attempts to minimize.

\subsection{Overview}
The \methodname~algorithm has two phases.
First, a \textit{proposal phase} (Section \ref{sec:proposal}) finds clusters of similar programs and uses these clusters to propose a set of candidate macros. 
Then, an \textit{integration phase} (Section \ref{sec:integration}) greedily iterates through a ranked list of these candidate macros and adds them to the library $\library$ whenever it would improve the objective function $\objective$. 
These phases can be alternated one after the other for multiple rounds, with the output of one phase treated as the input for the next (Fig.~\ref{fig:overview}).
By iterating this procedure for multiple rounds, increasingly complex macros can be found; as a macro discovered in round $t$ can use a previously-discovered macro from round $t-1$ in its definition (Fig.~\ref{fig:macro_method}).

\rev{Working with imperative programs that contain real-valued parameters presents unique challenges.
For instance, it is difficult to reason about valid line re-orderings of imperative programs when discovering macros and deciding when they can be applied. 
\methodname uses a sampling-based approach to discover macros by creating clusters of shapes with shared program structure (Section \ref{sec:form_cluster}) and a beam search procedure to decide how to apply discovered macros to existing programs (Section \ref{sec:best_prog}). 
Moreover, when dealing with real-valued parameters, it is challenging to find meaningful (non-spurious) parametric relationships, especially within a single program. 
To achieve generality, \methodname finds abstracted expressions that simultaneously describe multiple programs from a cluster of related shapes (Section \ref{sec:abs_prog}). 
}

Complete pseudocode for \methodname~is shown in Algorithm \ref{algo:mod}; Sections~\ref{sec:proposal} and \ref{sec:integration} explain this procedure in more detail.
As input, it takes in a starting library of functions $\library$, a dataset of imperative programs $\dataset$ and an objective function $\objective$ to be minimized.
Each element of $\dataset$ is a tuple $(\program,\orders_\program)$ containing program lines $\program$ and the set of valid orderings for those lines $\orders_\program$ (i.e. re-orderings of the lines which produce the correct output when executed).

\subsection{Initialization}
In our experiments, the library $\library$ is initialized with the 5 manually designed functions from the ShapeAssembly grammar. 
Then, starting with a collection of hierarchically-organized 3D cuboid structures from PartNet \cite{PartNet}, we use ShapeAssembly's data parsing algorithm to find program lines $\program$ which recreate each shape.
We then developed a procedure to determine the set of valid orderings $\orders_\program$ for that program (i.e. all orderings which produce the correct output geometry) to form our input dataset $\dataset$. 
Further details about the data parsing and valid ordering procedures can be found in the supplemental (Section A.3).

\definecolor{nice_gray}{rgb}{0.77,0.77,0.77}
\definecolor{nice_blue}{rgb}{0.62,0.85,0.89}
\definecolor{nice_pink}{rgb}{0.96,0.71,0.82}
\definecolor{nice_green}{rgb}{0.59,0.87,0.54}
\definecolor{nice_purple}{rgb}{0.77,0.68,0.83}

\newcommand{\colorbg}[2]{%
  \mbox{\vphantom{#2}\smash{\colorbox{#1}{#2}}}%
}

\subsection{Objective Function}
\label{sec:objective}

Our goal is to represent an entire dataset of programs compactly (removing free parameters) while also keeping the number of functions in the library small. 
Specifically, our objective is to minimize a weighted sum of the number of functions in $\library$ and the number of free parameters needed to represent programs in the dataset $\dataset$.
\rev{For ShapeAssembly,} free parameters can have multiple types $\Tau$:
Choice of function per line (\textbf{fn}), cuboid ID (\textbf{cid}), float/continuous (\textbf{f}), discrete (\textbf{d}), Boolean (\textbf{b}).
One may care about compressing these types differently, we allow each parameter type to be weighed differently in the objective defined as, 
\begin{equation*}
    \objective = \lambda_\mathbf{n} |\library| + \frac{1}{|\dataset|} \sum_{\tau \in \Tau} \lambda_{\mathbf{\tau}} |\tau(\programs^*(\dataset, \library))| 
    + \lambda_{\mathbf{\epsilon}} \epsilon(\tau, \dataset, \programs^*(\dataset, \library)) 
\end{equation*}
where $\programs^*(\dataset, \library)$ returns the best programs for $\dataset$ using the functions in $\library$ (Section~\ref{sec:best_prog}), $\tau(\programs)$ returns the set of all $\tau$-typed free parameters in the programs $\programs$, and $\epsilon(\tau, \dataset, \programs)$ returns the sum of errors in $\tau$-typed parameters incurred by using $\programs^*(\dataset, \library)$ in place of the original programs in $\dataset$.
The weights $\lambda_\mathbf{n}$, $\{\lambda_\tau | \tau \in \Tau\}$ and $\lambda_\epsilon$ can be adjusted to express preferences for the types of macros the algorithm aims to find.
In our experiments, we use $\lambda_\mathbf{n} = 1$, $\lambda_{\textbf{fn}} = 8$, $\lambda_{\textbf{cid}} = 8$, $\lambda_{\textbf{f}} = 1$, $\lambda_{\textbf{d}} = 0.5$, $\lambda_{\textbf{b}} = 0.25$, and $\lambda_\epsilon = 10$.  

\subsection{Finding the Best Program for a Given Library}
\label{sec:best_prog}

Calculating the value of $f$ over a dataset of shapes requires finding the program under $\library$ that minimizes the objective function for each program $(\program, \orders_\program) \in \dataset$. 
As $\orders_\program$ is a collection of valid orderings of the program lines $\program$, we solve this problem by finding the best scoring program under $\library$ for every $\order \in \orders_\program$.
Combining an ordering $\order$ with program lines $\program$ produces a program expressed in terms of base library functions $\program_\order$.
We then want to find the best program, $\program^*$, that uses the functions in $\library$ (including macros, if $\library$ contains them) to recreate $\program_\order$ while minimizing $\objective$.
We implement this procedure with a beam search that iteratively builds partial programs in the beam by adding calls to functions from $\library$ whose expansions cover lines in $\program_\order$. 
For a function expansion to \emph{cover} a sequence of program lines, the expansion must match those lines on command type, the values of the discrete / Boolean parameters must match exactly, and the continuous parameters must differ by an amount no greater than $\epsilon$. 
\rev{We set $\epsilon= 0.05$, finding that larger values lead to abstracted programs with degenerate geometry.}
We rank partial programs in the beam by their objective value, normalized by the number of lines in $\program_\order$ it is covering.
This search runs until all programs in the beam have no more lines in $\program_\order$ to cover; the program with lowest objective value is returned as the best program $\program^*$.
In the case of ties, we choose the program with the most canonical ordering, as explained in the supplemental material.
In our implementation, we use a beam width of 10. 
\rev{Other search strategies could be applied here; we chose beam search as it was relatively fast and found good solutions.}

\section{Proposal Phase}
\label{sec:proposal}

\begin{figure*}[t!]
  \includegraphics[width=\linewidth]{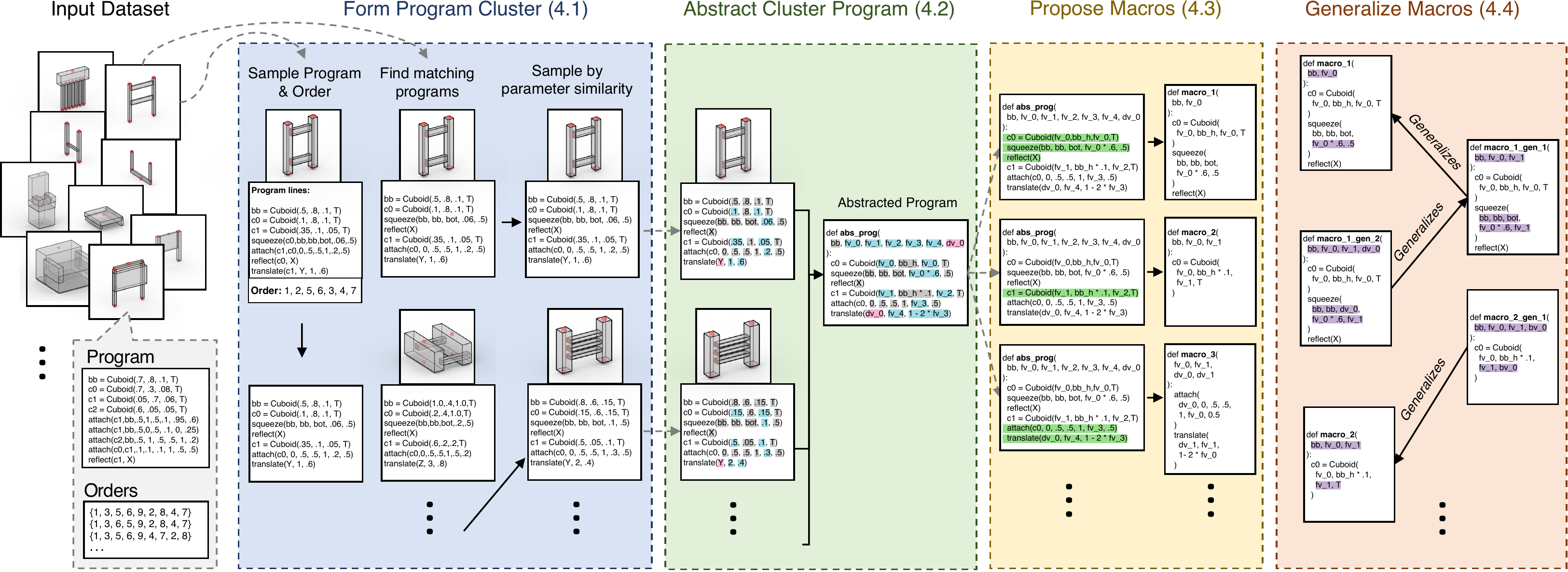}
\caption{\methodname's proposal phase, which proposes candidate macros to be added into \library.
Each round of this phase begins by identifying a cluster of structurally-identical programs with similar parameter values within the input dataset (Section~\ref{sec:form_cluster}).
It then finds a single abstracted program which subsumes most or all of the programs in this cluster (Section~\ref{sec:abs_prog});
here, gray parameter values are abstracted as constants, blue ones as continuous free variables, and pink ones as discrete free variables.
Subsequences of lines in this abstracted program (shown in green) are isolated to form potential macros which could be used to re-write the program (Section~\ref{sec:cand_funcs}).
Finally, this set of candidate macros is expanded by including generalizations of the initial set (Section~\ref{sec:gen_prog}); purple lines show lines that are generalized.
Best viewed on a high-resolution screen.}
\label{fig:proposal_phase}
\end{figure*}

The goal of \methodname's proposal phase is to construct a set of candidate macros which might be useful for compressing the dataset of shape programs $\dataset$.
A schematic overview of the proposal phase is shown in Figure ~\ref{fig:proposal_phase}.
In each proposal round, the algorithm first forms a cluster of similar programs sampled from $\dataset$ (Section \ref{sec:form_cluster}). 
Then, using the functions of $\library$, it finds an abstracted program that explains the majority of examples in the cluster while trying to remove free parameters whenever possible (Section \ref{sec:abs_prog}). 
It converts this abstracted program into a set of candidate macros (Section \ref{sec:cand_funcs}) and finds potential generalizations of these macros (Section \ref{sec:gen_prog}).
This process is repeated for \emph{num\_proposal\_steps} (we use 10000) to build up a large collection of candidate macros.

\subsection{Form a Program Cluster}
\label{sec:form_cluster}

The goal of the cluster formation step is to find a set of programs from $\dataset$ that can be represented by a single abstracted program, i.e. a program with free variables.
The blue box in Fig.~\ref{fig:proposal_phase} illustrates the procedure.
The algorithm first randomly samples a program $\program$ from $\dataset$ and then randomly samples an order $\order$ from the possible valid orderings in $\orders_\program$ (Algorithm~\ref{algo:mod}, line 4). 
It then finds the set of programs $\programs_\textbf{matches}$ from $\dataset$ that \emph{structurally match} $\program$ and also have $\order$ as one of their valid orderings in (Algorithm~\ref{algo:mod}, line 5).
In ShapeAssembly, two programs structurally match if they use the same set of commands which refer to the same cuboid IDs (though their other parameters may vary).

For each program in $\programs_\textbf{matches}$, we record the norm $n$ of the difference of its continuous parameters compared with those in $\program$. 
We then form a probability distribution over $\programs_\textbf{matches}$, where each program is given a weight proportional to $1 - \frac{n}{n*}$, where $n*$ was the maximum observed $n$.
\rev{Taking the parameter distance between programs into account results in clusters that are more semantically consistent, which increases the likelihood the abstracted program we produce can discover meaningful parametric relationships.}
Finally, we sample $k$ programs from $\programs_\textbf{matches}$ using this probability distribution in order to form $\programs_\textbf{cluster}$ (line 6).
We set $k$ to 20.

\subsection{Find Abstracted Program for Cluster}
\label{sec:abs_prog}

Given the cluster of programs $\programs_\textbf{cluster}$ identified in the previous section, the next step is to use the library of functions $\library$ to find the most compact program (fewest free parameters) that can represent the majority of programs in $\programs_\textbf{cluster}$ (Algorithm~\ref{algo:mod}, line 7). 
By construction, the sequence of functions and cuboid IDs is the same across all programs in $\programs_\textbf{cluster}$. 
To build up the abstracted program $\program_\textbf{abs}$, the algorithm uses a similar procedure to the best-program-finding routine in Section~\ref{sec:best_prog}: covering each line in the cluster by choosing functions from $\library$.  
However, instead of using a beam search to find the sequence of functions, here we employ a greedy strategy. 
We create a preference ordering over the functions of $\library$ based on how many free parameters each function constrains (weighted by their respective $\lambda_\tau$ weights).
Then, whenever we need to pick a function, we step through this ordering, until we find a function that is able to match the parameters of at least $p = 70\%$ of the next lines from $\programs_\textbf{cluster}$.

For each function added to the abstracted program, we iterate through its parameter slots to see if we can remove more degrees of freedom. 
For discrete parameters, a constant can be used, a previously defined parameter can be used, or a new free parameter can be declared. 
For continuous parameters, a constant can be used, an expression over previously defined parameters can be used, or a new free parameter can be declared. 
The details of this logic can be found in the supplemental material (Section B.1).
In all cases, the value chosen for each parameter must still be valid for at least $p$ percent of programs in $\programs_\textbf{cluster}$.
This process iterates until there are no remaining uncovered lines in the programs of $\programs_\textbf{cluster}$.
At this point, $\program_\textbf{abs}$ is complete.
The green box in Fig.~\ref{fig:proposal_phase} shows an example of finding a single abstracted program for two base programs.

\begin{figure*}[t!]
  \includegraphics[width=\linewidth]{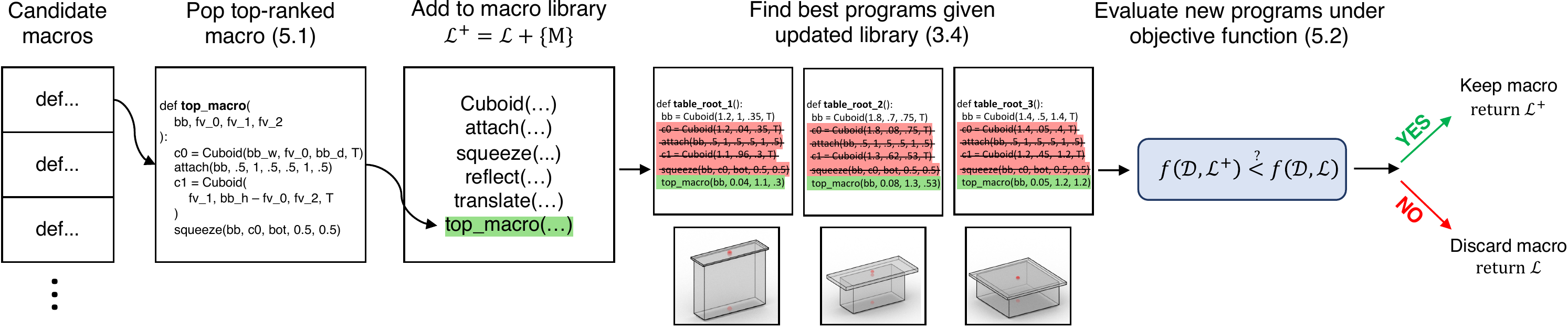}
\caption{
\methodname's integration phase, which chooses which candidate macros to add to the DSL library $\library$.
On each round of this phase, the algorithm heuristically ranks candidate macros based on which are likely to improve program compression, adds the top-ranked macro to the library, then finds the best refactored program for each program in the input dataset $\dataset$ under this new library. If this refactoring lowers the objective value $\objective(\dataset,\library)$, then the macro is kept in the library; otherwise, it is discarded.
}
\label{fig:integration_phase}
\end{figure*}

\subsection{Proposing Candidate Macros}
\label{sec:cand_funcs}

The abstracted program $\program_\textbf{abs}$ found in the previous step represents multiple shape programs from our dataset (via leaving some of its parameters as free parameters).
Thus, its function body likely contains re-usable shape programming patterns---in other words, it is a good source of potential macros $\macros$ (Algorithm~\ref{algo:mod}, line 8).
In this next step, the algorithm iterates through the lines of $\program_\textbf{abs}$ and finds all line sequences that could be turned into a valid macro (yellow box in Fig.~\ref{fig:proposal_phase}). 
A valid macro $\macro$ is a sequence of program lines that simplifies the program, i.e. it must remove some degree of freedom from the program lines it aims to cover. 
For both computational efficiency, and to encourage the creation of more meaningful macros, we impose some additional restrictions on the definition of a valid macro; see the supplemental material (Section B.2).
For each created candidate macro, we record what cluster it was found in and the lines of the cluster it covered, in order to calculate frequency statistics used later in the integration phase (Section~\ref{sec:integration}).

\subsection{Generalizing Macros}
\label{sec:gen_prog} 

As $\program_\textbf{abs}$ is designed to maximally condense all of the programs in $\programs_\textbf{cluster}$, the generated candidate macro operators $\macros$ may be somewhat overly-specific to the subset of programs in $\programs_\textbf{cluster}$.
Furthermore, $\macros$ may also contain some very similar macros \rev{that are treated as distinct}.
To get around these issues, the proposal phase concludes with a generalization step, where for each discovered candidate macro, we also find all generalizing macros that are within $n$ program edits (Algorithm~\ref{algo:mod}, line 9). 
\rev{We set $n = 2$ due to running time constraints; in principle, higher values of $n$ will lead to better solutions.}
For a given macro $\macro$, another macro $\macro'$ is defined to be generalizing if for every parameterization of $\macro$, $\macro'$could be parameterized to produce the same output.
From this generalization procedure we form a graph where each node is a macro and edges between two nodes indicates a generalizing relationship (orange box in Fig.~\ref{fig:proposal_phase}). 
This graph is used to update frequency statistics (in that generalizing macros also cover all lines covered by macros they generalize) which influences the candidate macro ranking logic used by the integration phase (Section~\ref{sec:integration}).

\section{Integration phase}
\label{sec:integration}

Given candidate macros from the proposal phase, the integration phase chooses which macros to add to the library $\library$ in order to minimize its objective function $\objective$.
Figure~\ref{fig:integration_phase} shows an overview.
Solving such a subset selection problem optimally is NP-hard, so this phase instead employees a greedy approximation.
It iterates through the candidate macro operators, on each iteration taking the highest ranked macro based on expected improvement to $\objective$ (Section \ref{sec:rank_cand_abs}).
It then decides whether to add the macro into the library $\library$ by evaluating its effect on the objective function (Section \ref{sec:eval_cand_abs}).

\subsection{Ranking Candidate Macros}
\label{sec:rank_cand_abs}

The proposal phase can generate tens of thousands of candidate macros; it is computationally intractable to consider all of them.
To prioritize which candidate macros to consider within a finite time budget, the algorithm employs a heuristic ranking scheme (Algorithm~\ref{algo:mod}, line 13).
The rank of a candidate macro $\macro$ is based on an estimate of how much using $\macro$ would improve the score of the objective function.
The ranking scheme first calculates the \emph{gain} of the macro over the functions already in $\library$. 
The gain $g$ of a macro $\macro$ is the weighted sum of the number of free parameters (weighted by their respective $\lambda_\tau$ weights) that would be removed each time $\macro$ were used in a program instead of the lowest-cost sequence of functions currently in $\library$ that is equivalent to or generalizes $\macro$.
Then our ranking scheme calculates the percentage of shapes $p$ that produced $\macro$ as a candidate macro during the proposal phase. 
The ranking score of $\macro$ is then simply $p \cdot g$.
This score is a simple estimate of the effect on the actual objective value $\objective(\dataset, \library + \{\macro\})$ that does not require the expensive step of finding the best programs for the whole dataset.

\subsection{Evaluating \& Selecting Candidate Macros}
\label{sec:eval_cand_abs}

Given a candidate macro operator $\macro$, the next step is to see if adding it to $\library$ would actually improve the value of the objective function $\objective$. 
For this, we define a function \texttt{optimize} which takes in $\objective$, the current library $\library$, a modified version of the library $\library^+$ , and a subset of programs from the dataset $\tilde{\dataset} \subset \dataset$.
It returns whichever version of the library has the lower objective value, i.e. $\text{argmin}(f(\tilde{\dataset}, \library^+) < f(\tilde{\dataset}, \library))$.
Using a subsample $\tilde{\dataset}$ of the full dataset reduces computation time, i.e. we are using an unbiased estimator of the true objective value for the dataset.

The algorithm first calls \texttt{optimize} with a modified library where $\macro$ is added to $\library$ (line 14). 
If this leads to a library change, then it continues to the next candidate macro operator (lines 15-16).
If $\library$ remains unchanged, it checks if any of the functions currently in $\library$ are used significantly less in finding the best programs over $\tilde{\dataset}$ when the modified library version is used (line 17). 
If the set of functions in $\library$ whose frequency decreased significantly, $\macro_\textbf{infreq}$, is not empty, then it runs \texttt{optimize} once again with a modified version of the library that includes $\macro$ but removes all elements of $\macro_\textbf{infreq}$ (lines 18-20). 
This step allows the algorithm to avoid a local minima where $\macro$ would not be added to $\library$, even if it could ultimately improve $\objective$, because similar macros to $\macro$ had been added to $\library$ earlier. 
If this step changes the library, then $\library$ has been updated to include $\macro$, but it does not include any of the functions in $\macro_\textbf{infreq}$. 
Thus, the algorithm attempts to add each $\macro \in \macro_\textbf{infreq}$ back into $\library$, by once again calling \texttt{optimize} and keeping the library version with the better score (lines 23-24).
Finally, after evaluating \emph{num\_integration\_steps}=20 macros, the algorithm checks if $\objective$ can be improved by removing any of the functions in $\library$ (lines 25-26). This can be beneficial, for instance, when a macro discovered in an early round becomes a sub-routine of a macro discovered in a later round, and therefore appears less frequently (or not at all) in $\programs^*(\dataset, \library)$.

\subsection{Removing Bad Program Orders}
\label{sec:filter_bad_orders}

When $\library$ is composed of only original library functions, any valid ordering in $\orders_\program$ for $\program$ will lead to a program that produces the same score under $\objective$. 
As macros are added into $\library$, using different line orderings in $\orders_\program$ may result in different scores under $\objective$ (as some line orders will prohibit certain macros from being applied). 
As such, after each integration round, the algorithm removes any orders from $\orders_\program$ that lead to objective function scores that are significantly worse (using a threshold of $\tau_\order = 1$) then the score produced by the order, $\order^*$; the order that leads to the best objective function score for $\program$ (Algorithm~\ref{algo:mod}, line 27). 
The following proposal rounds will then only be able to use orderings that have not been filtered out of $\dataset$. 
\rev{Keeping the orderings that perform best during the preceding integration phase produces more accurate heuristic rankings of  macros from the proposal phase (Section \ref{sec:rank_cand_abs}). We found this encouraged the discovery of complex macros, e.g. without this step, the ‘four leg base’ macro was not discovered.}

\section{Results and Evaluation}
\label{sec:results}

We experimentally evaluate \methodname's effectiveness at compressing shape programs and at supporting downstream tasks.
Our experiments use three categories of manufactured shapes (Chairs, Tables, Storage) from CAD models in PartNet.
We use the same data parsing procedure as described in the original ShapeAssembly paper~\cite{jones2020shapeAssembly} to produce 3836 Chair programs, 6536 Table programs, and 1551 Storage programs.
In Section \ref{sec:res_disc}, we examine the properties of \methodname's discovered macros on dataset compression.
In Section \ref{sec:res_gen}, we show that using these macros improves the performance of generative models of 3D shape structures.
In Section \ref{sec:res_infer}, we demonstrate that macros aid in visual program induction tasks. 
And finally, in Section \ref{sec:res_edit}, we report the results of a user study comparing performance on goal-directed shape editing tasks with and without discovered macros.


\subsection{Discovered Macros}

\begin{table}[t!]
    \centering    
    \setlength{\tabcolsep}{2pt}
    \footnotesize
    \caption{We measure how well different libraries can compress a dataset of shape programs (metric details in Section \ref{sec:res_disc}). 
    For all compression metrics, lower values are better, as our goal is to find a small collection of functions that remove many degrees of freedom from the underlying shape programs. \methodname operates by attempting to minimize \objective, and we show that it does in fact improve \objective~ compared to the No Macros version. 
    }
    \begin{tabular}{@{}llcccccc@{}}
        \toprule
        \textbf{Category} & \textbf{Method} 
        & \textbf{\objective} 
        & \textbf{$|\library|$} 
        & \textbf{fn$(\programs^*)$} 
        & \textbf{d$(\programs^*)$} 
        & \textbf{f$(\programs^*)$} 
        & \textbf{b$(\programs^*)$} 
        \\
        \midrule
        \multirow{3}{*}{\emph{Chair}}
        & No Macros & 411 & \textbf{5} & 29.8 & 17.8  & 84.4 & 11.3  \\
        & Baseline Macros & 312 & 36 & 21.7 & 7.0 & 80.2 & \textbf{4.2} \\
        & \methodname & \textbf{260} & 17 & \textbf{21.0} & \textbf{6.4} & \textbf{58.1} & 8.6 \\
        \midrule
        \multirow{3}{*}{\emph{Table}}
        & No Macros & 356 & \textbf{5} & 25.6 & 16.3 & 70.7 & 9.6 \\
        & Baseline Macros & 263 & 36 & 18.0 & 6.4 & 65.8 & \textbf{3.2} \\
        & \methodname & \textbf{214} & 15 & \textbf{17.4} & \textbf{5.1} & \textbf{48.7} & 5.6 \\
        \midrule
        \multirow{3}{*}{\emph{Storage}}
        &  No Macros & 453 & \textbf{5} & 30.4 & 21.6 & 92.2 & 11.7  \\
        & Baseline Macros & 314 & 48 & \textbf{18.4} & \textbf{7.6} & 88.45 & \textbf{2.65}  \\
        & \methodname & \textbf{283} & 17 & 21.1 & \textbf{7.6} & \textbf{68.9} & 4.0 \\
        \bottomrule
    \end{tabular}
    \label{tab:prog_stats}
\end{table}

\label{sec:res_disc}
\rev{For each shape category, we run \methodname until \objective stops decreasing (5 rounds in all cases) to discover a small set of macro operators.}
Instead of applying \methodname directly on hierarchical programs, we form ~\dataset~ by decomposing each ShapeAssembly program into a collection of non-hierarchical sub-programs (e.g., a single Chair might contribute one program for its back sub-part and one program for its base sub-part). 
We implement \methodname in Python and run the algorithm on a computer with an Intel i9-9900K CPU, which takes 5 hours for Chairs, 12 hours for Storage, and 19 hours for Tables.  

\begin{figure*}[t!]
  \includegraphics[width=\linewidth]{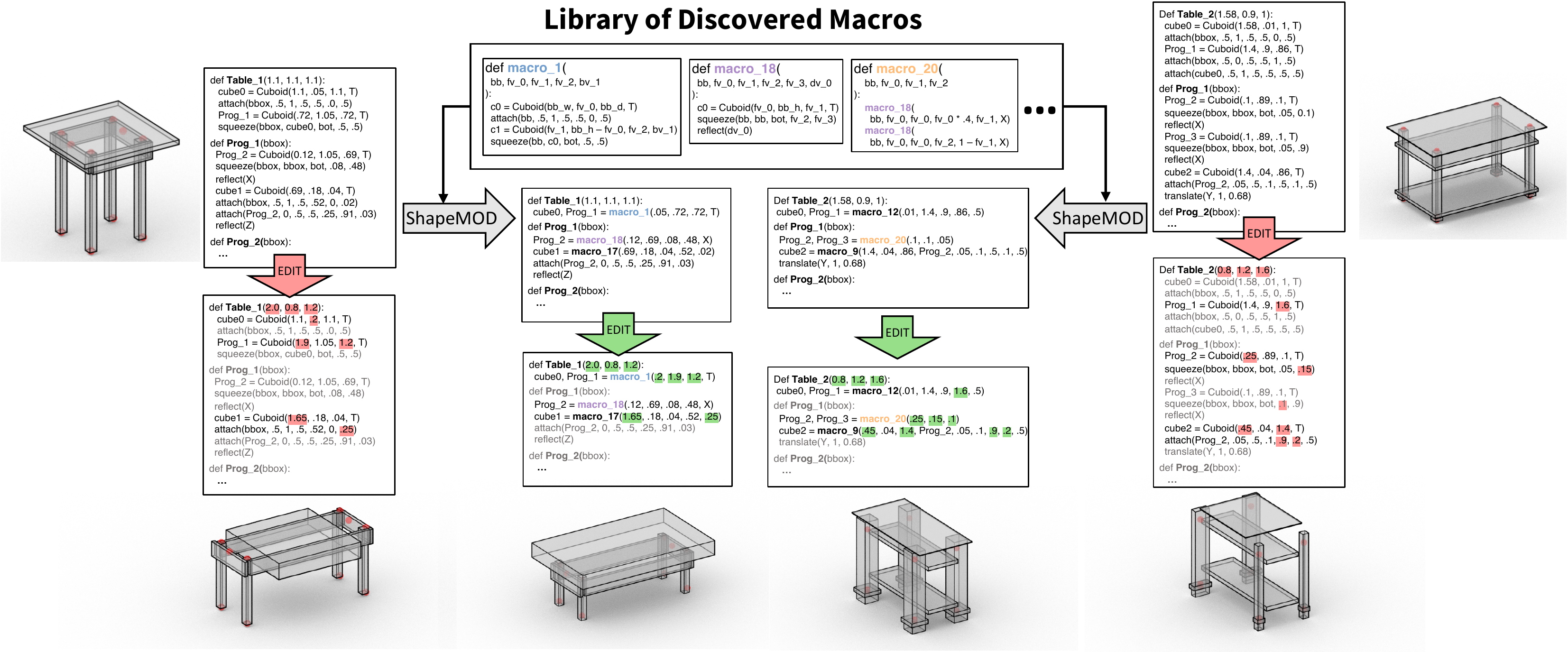}
\caption{
We show some macros (top-middle) that \methodname discovered when run on the Table dataset, and program refactors that use these macros to significantly compress the number of exposed free parameters (\methodname arrows from outside to inside).
We show program edits (down arrows) of corresponding parameters in both programs with macros (green) and without macros (red). 
The discovered macros capture parametric relationships that better preserve shape plausibility under manipulation; for example, all chair legs remain the same size in the third column (macros), while the shape in the fourth column (no macros) becomes disconnected and physically implausible .  
}
\label{fig:qual_funcs}
\end{figure*}

Fig.~\ref{fig:qual_funcs} shows examples of some of the macros discovered for Tables; see the supplemental material for complete discovered libraries for all shape categories (Section F).
These macros are used by multiple shape programs in our dataset, explaining common patterns and shortening programs that use them.
They also better facilitate editing: making edits to a few parameters in macro-refactored programs tends to produce more plausible shape variations than edits to the corresponding parameters of the macro-free program. 
For instance, discovered macro\_1 introduces a relationship that the heights of the table base and the table top should sum to the height of the table bounding box. 
Without this macro, edits to base ShapeAssembly functions can easily cause the table top to overlap and intersect parts of the table base in an implausible manner (left side of figure).

We compare the library of functions generated by our \methodname procedure to two baselines:
\begin{enumerate}[(i)]
\denselist
    \item \textbf{No Macros}:  The base library of functions from ShapeAssembly that is used to initialize our \methodname procedure.
    \item \textbf{Baseline Macros}: A naive single-pass approach for macro discovery that creates macros out of the most common structural sequences present in the dataset and replaces parameters with constants whenever a high percentage of its parameterizations share similar values. See Appendix \ref{sec:baseline_method} for details.
\denselist
\end{enumerate}
Table \ref{tab:prog_stats} compares these baselines to \methodname's discovered language on the task of compressing a dataset of 3D Shape programs. We consider the following metrics:
\begin{itemize}
\denselist
    \item Value of \methodname's objective function (\objective)
    \item Number of functions in library ($|\library|$)
    \item Number lines in the best programs ($\textbf{fn}(\programs^*)$)
    \item Number of discrete parameters in the best programs ($\textbf{d}(\programs^*)$
    \item Number of continuous parameters in the best programs ($\textbf{f}(\programs^*)$)
    \item Number of Boolean parameters in the best programs ($\textbf{b}(\programs^*)$)
\denselist
\end{itemize}
By adding only a handful of macros to the language, \methodname significantly compresses programs in terms of number of lines and number of free parameters.
For instance, the 12 Chair macros discovered remove 30\% of program lines, 64\% of the discrete parameters, and 30\% of the continuous parameters needed to represent the same dataset without macros. In total, these macro functions are able to decrease the value of the objective function we aim to minimize by 37\%.
Moreover, \methodname is able to compress programs to a greater degree than the baseline approach, especially for continuous parameters, while using half as many (or fewer) new macros.
\rev{We examine the effects of different design decisions on the convergence properties of \methodname with an ablation experiment, described in Appendix \ref{sec:convergence}.}

\begin{figure}
  \includegraphics[width=0.9\linewidth]{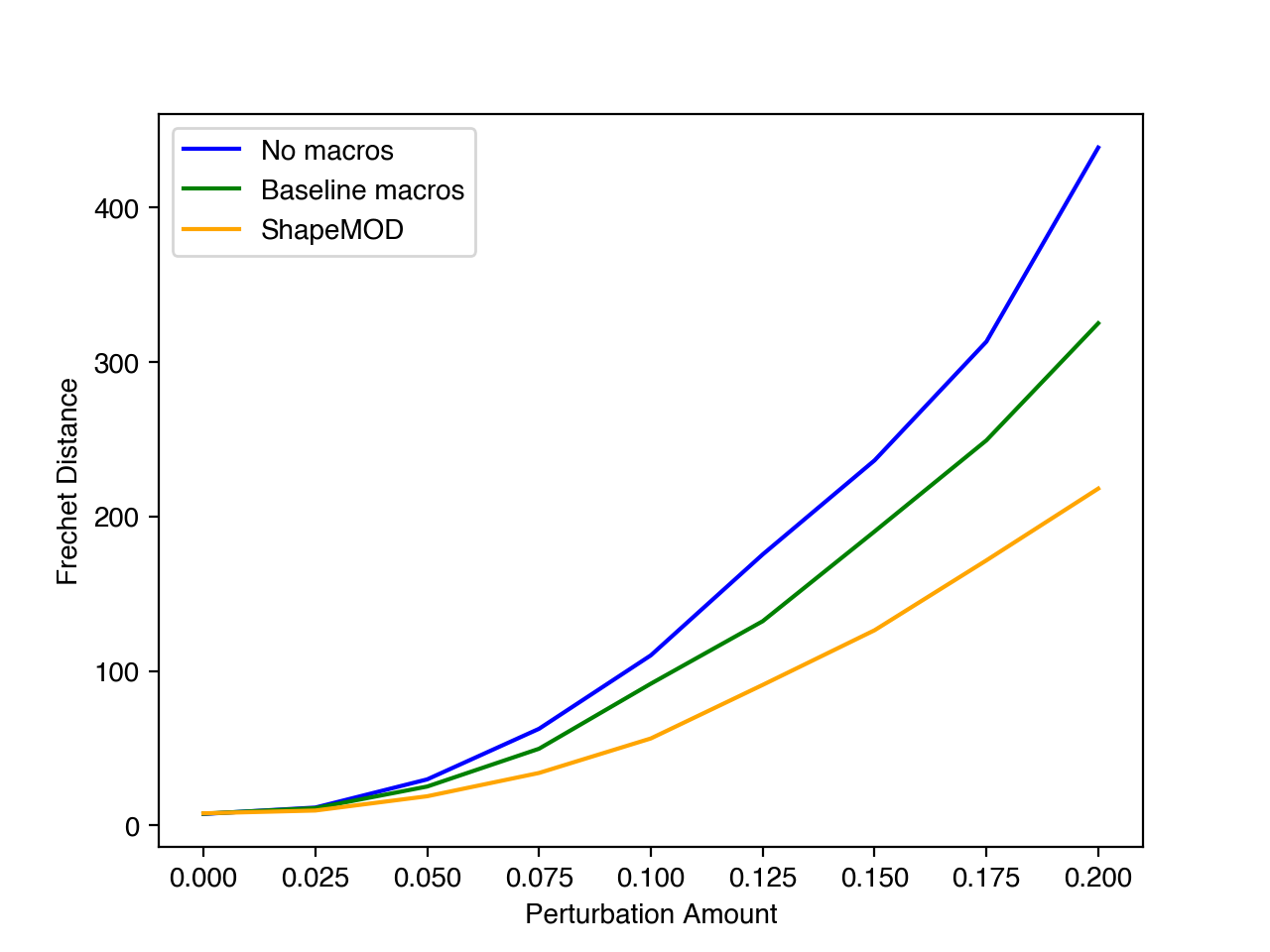}
\caption{
We measure distributional similarity (Frechet Distance) between a set of reference chairs and a set of chair programs subjected to perturbations.
We simulate perturbations by adding noise from a normal distribution (x-axis is $\sigma$) to continuous parameters in the programs. 
Programs with \methodname macros retain more similarity under larger perturbations, suggesting the macros remove degrees of freedom that permit shapes to move outside of their original distribution. 
}
\label{fig:perturb}
\end{figure}

\begin{figure*}[t!]
    \centering
    \setlength{\tabcolsep}{1pt}
    \begin{tabular}{ccccccccc}
        \raisebox{2.0em}{\textbf{Geom NN}  } &
        \includegraphics[{width=.1\linewidth}]{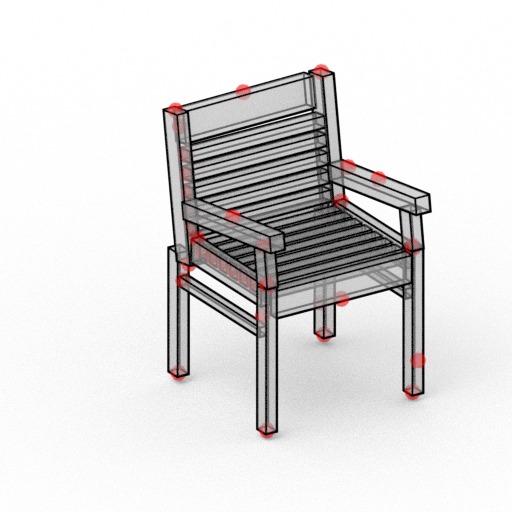} &
        \includegraphics[{width=.1\linewidth}]{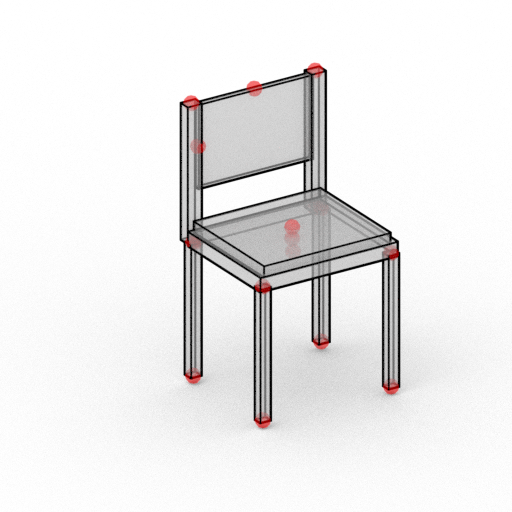} &
        \includegraphics[{width=.1\linewidth}]{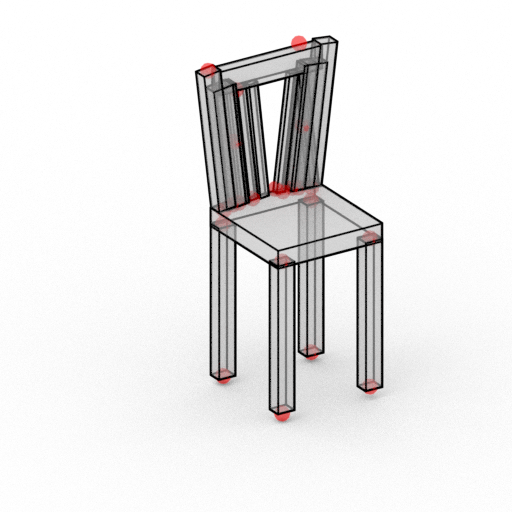} &
        \includegraphics[{width=.1\linewidth}]{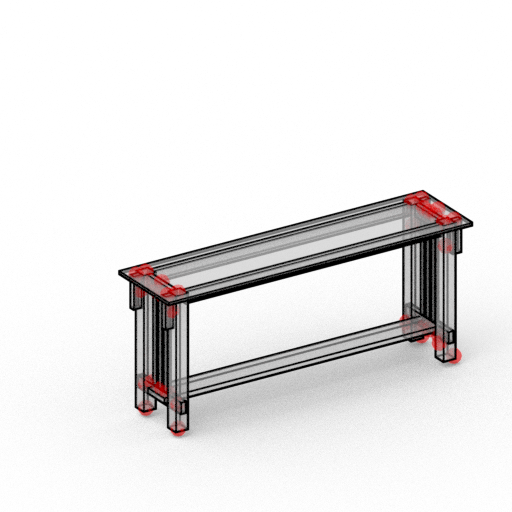} &
        \includegraphics[{width=.1\linewidth}]{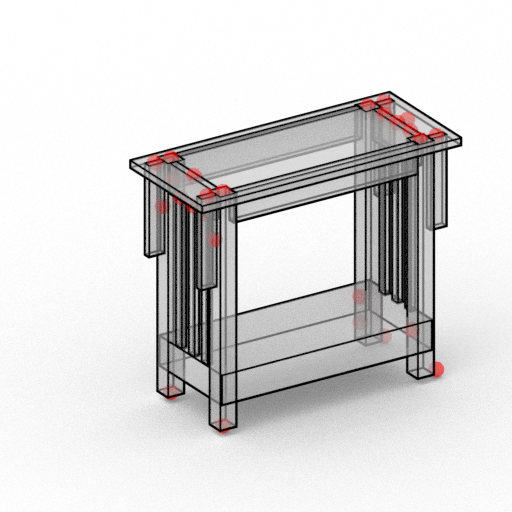} &
        \includegraphics[{width=.1\linewidth}]{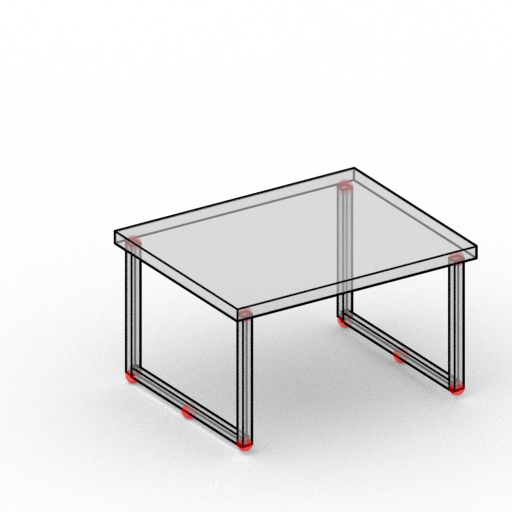} &
        \includegraphics[{width=.1\linewidth}]{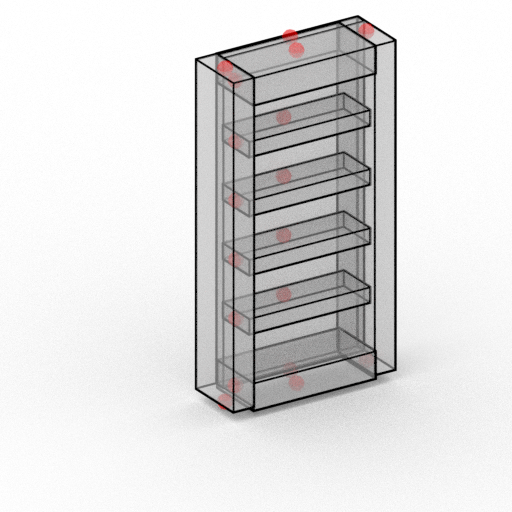} &
        \includegraphics[{width=.1\linewidth}]{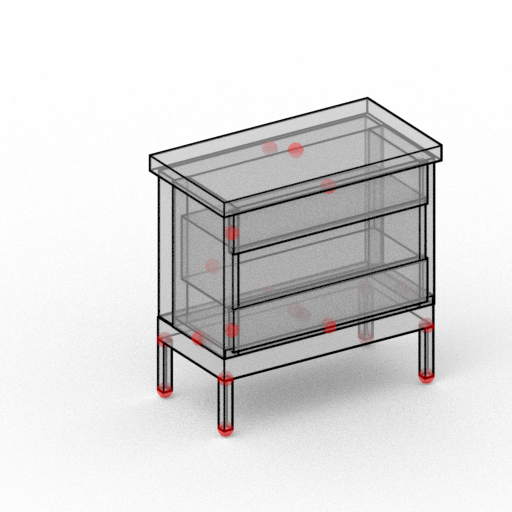} \\
        
        \raisebox{2.0em}{\textbf{\methodname }} &
        \includegraphics[{width=.1\linewidth}]{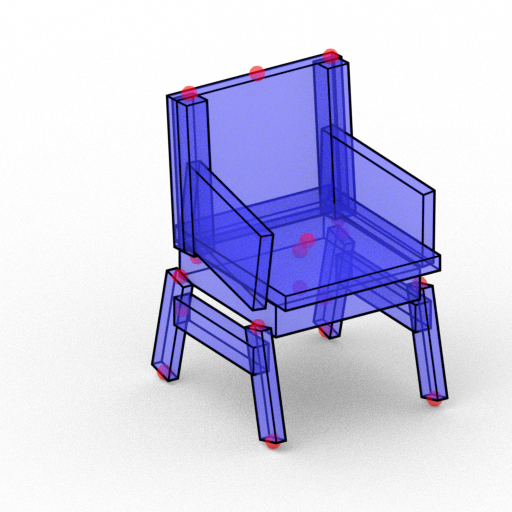} &
        \includegraphics[{width=.1\linewidth}]{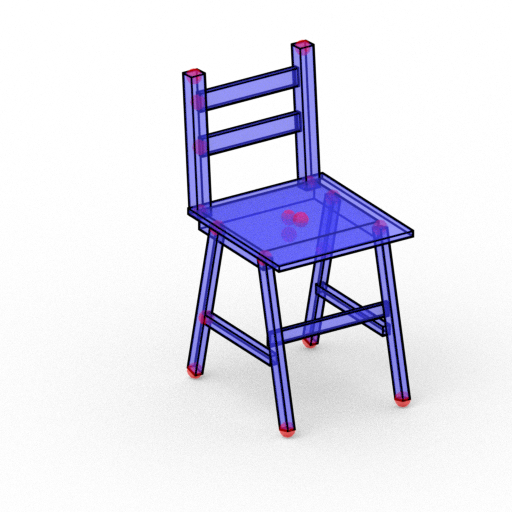} &
        \includegraphics[{width=.1\linewidth}]{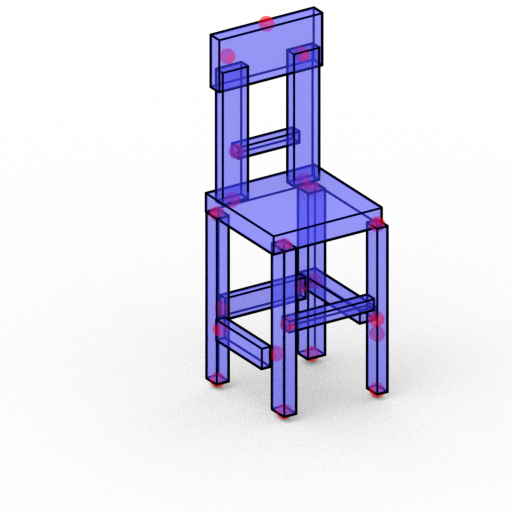} &
        \includegraphics[{width=.1\linewidth}]{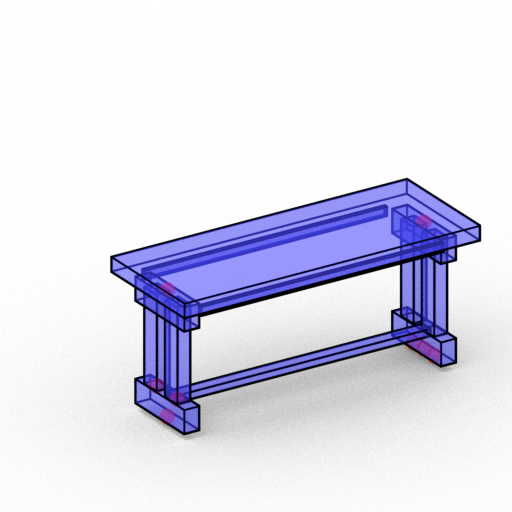} &
        \includegraphics[{width=.1\linewidth}]{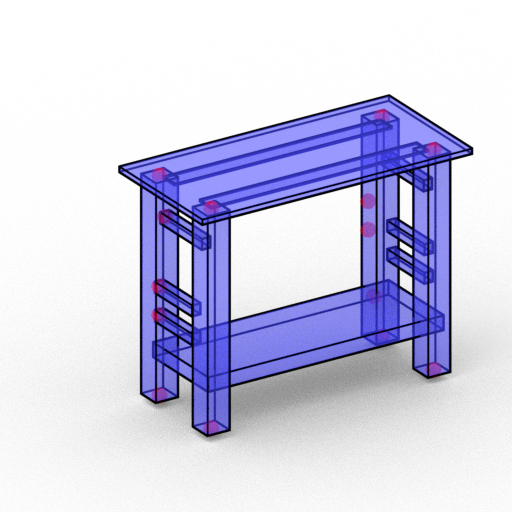} &
        \includegraphics[{width=.1\linewidth}]{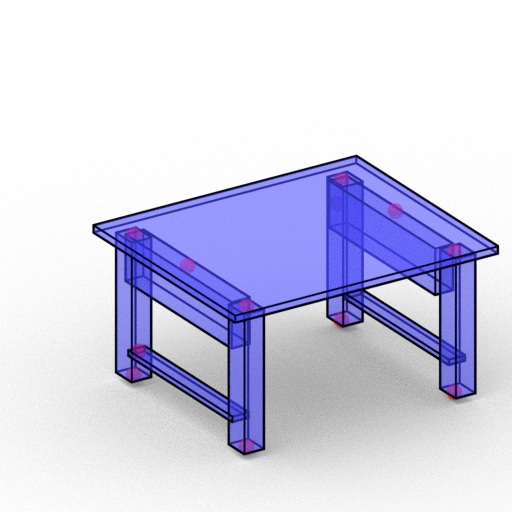} &
        \includegraphics[{width=.1\linewidth}]{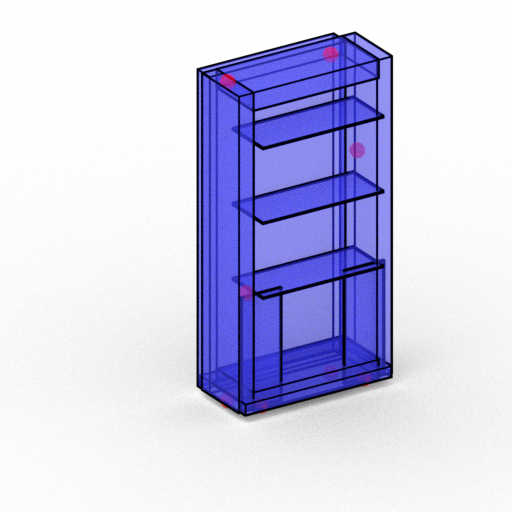} &
        \includegraphics[{width=.1\linewidth}]{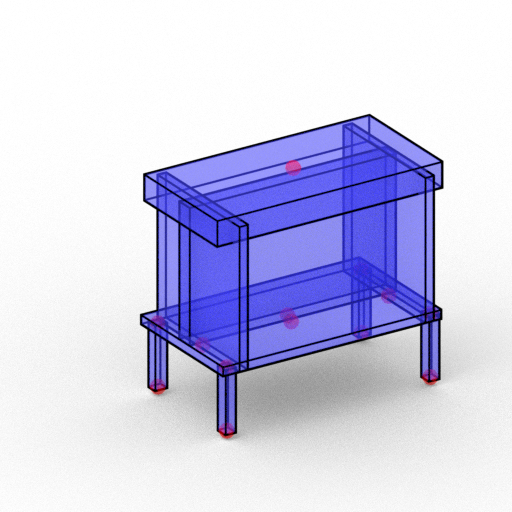} \\

        \raisebox{2.0em}{\textbf{Prog NN}  } &
        \includegraphics[{width=.1\linewidth}]{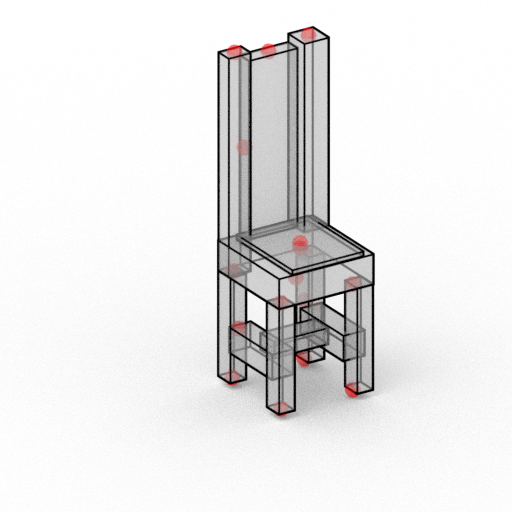} &
        \includegraphics[{width=.1\linewidth}]{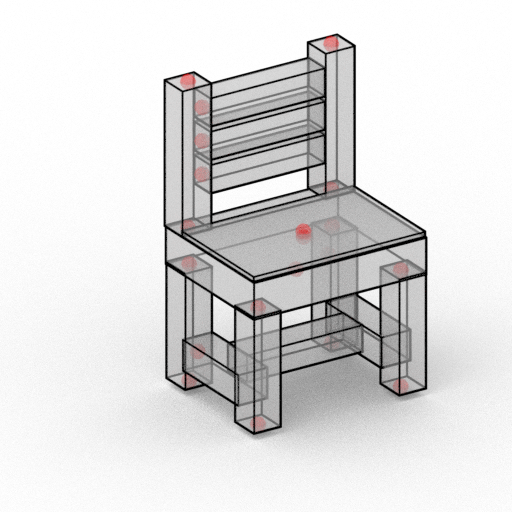} &
        \includegraphics[{width=.1\linewidth}]{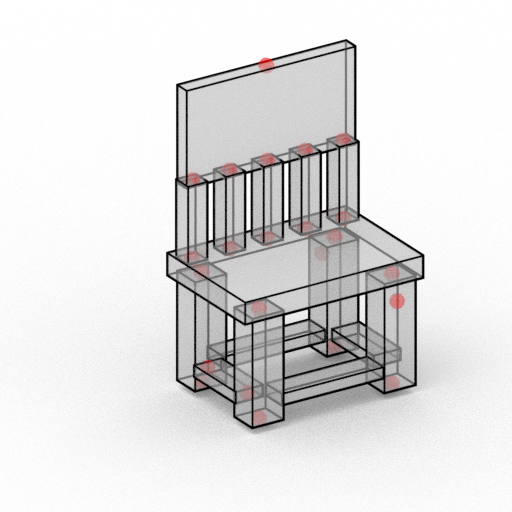} &
        \includegraphics[{width=.1\linewidth}]{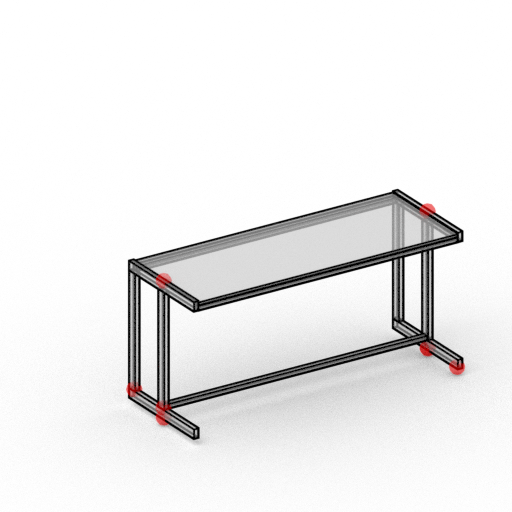} &
        \includegraphics[{width=.1\linewidth}]{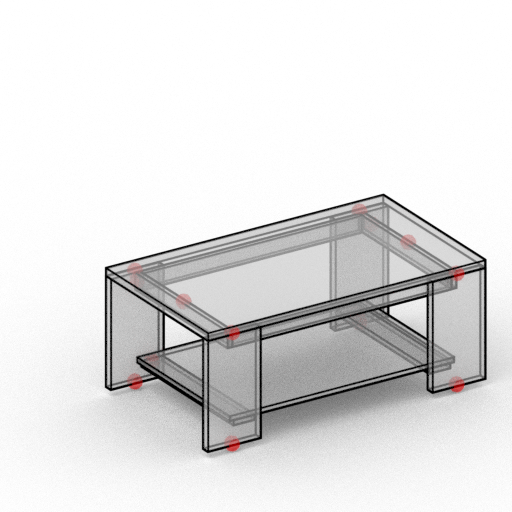} &
        \includegraphics[{width=.1\linewidth}]{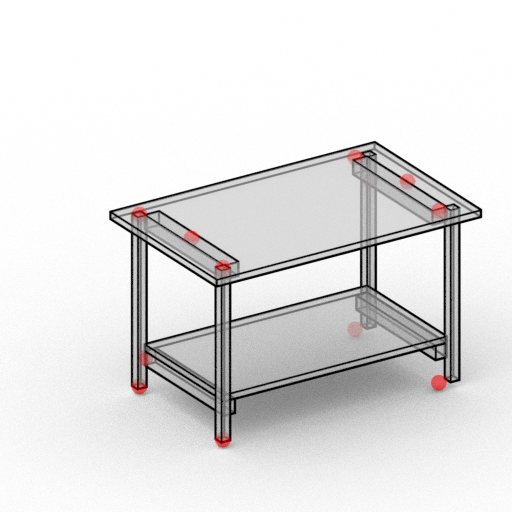} &
        \includegraphics[{width=.1\linewidth}]{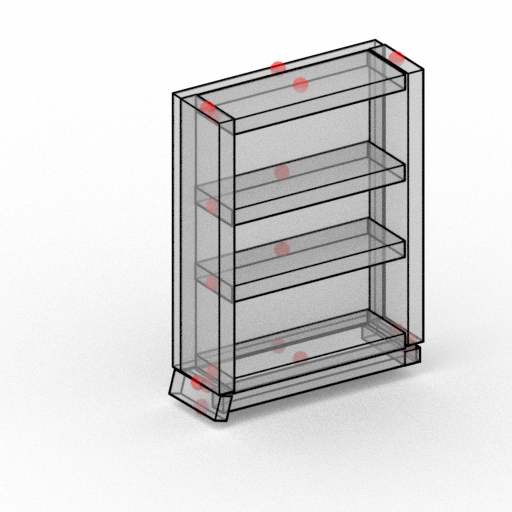} &
        \includegraphics[{width=.1\linewidth}]{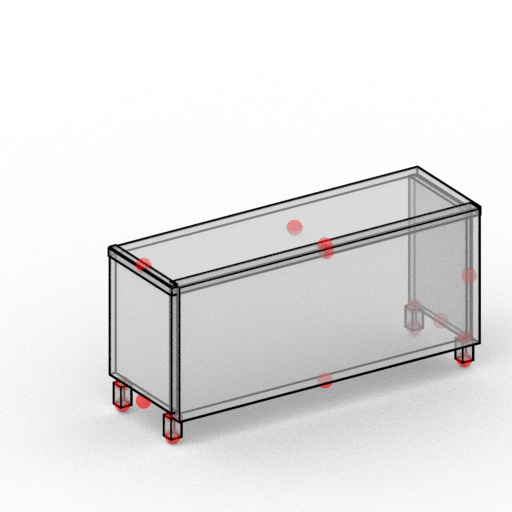} \\
       
    \end{tabular}
    \caption{
    Some example outputs of generative models trained to produce ShapeAssembly programs expressed with macros discovered by \methodname, along with their training set nearest neighbors (NN) by geometric and program similarity.
    Each cuboid represents a part proxy bounding volume. Structures are formed through attaching parts to one another (red dots).
    The generative models produce a variety of plausible structures without memorizing their training data.
    All corresponding programs can be found in supplemental material.
    }
    \label{fig:gen_nn}
\end{figure*}

The examples shown in Fig.~\ref{fig:qual_funcs} suggest that programs refactored using \methodname macros produce more plausible shapes under variations of their free parameters.
We ran an experiment to quantify this behavior.
Given a set of ground-truth Chair programs, we run \methodname and our baseline macros procedure on them to create a set of macro-refactored programs.
We then perturb the free parameters of both macro refactored and no macro programs by increasingly large perturbations, and we check how distributionally similar the outputs of the perturbed programs are to a held-out validation set of Chair shapes using Frechet Distance~\cite{FrechetInceptionDistance} in the feature space of a PointNet classifier pre-trained on ShapeNet \cite{qi2017pointnet, chang2015shapenet}.
Figure~\ref{fig:perturb} plots this distance against the magnitude of parameter perturbation.
Frechet Distance increases more slowly for programs that use macros, and increases the slowest for macros found using \methodname.
This indicates that the modes of variation in programs expressed with our method's macros are better at producing plausible output shapes that stay within the distribution that the collection of input programs originally came from.
In Section~\ref{sec:res_edit}, we conduct a shape-editing user study to further validate this behavior.

\subsection{Generating 3D Shapes}
\label{sec:res_gen}

We are interested in how well \methodname's discovered macros support the downstream task of generative shape modeling.
Our hypothesis is that using macros will restrict the output space of a program-generating model, making it harder to output `garbage' shapes.
To test this hypothesis, we train generative models on programs with and without \methodname macros.

For our generative model, we use the variational autoencoder architecture from ShapeAssembly~\cite{jones2020shapeAssembly}, modified slightly to support programs that use an arbitrary number of functions as opposed to a fixed, predefined set (see Appendix \ref{sec:modeling} for details). 
We train each model for 5000 epochs with a learning rate of $2e^{-4}$ and a batch size of 64. At the end of training, we choose the model from whichever training epoch produced the lowest Frechet Distance \cite{FrechetInceptionDistance} to the training set; we report all other metrics on a held out set.
Training was done on a computer with a GeForce RTX 2080 Ti GPU with an Intel i9-9900K CPU, consumed 2GB of GPU memory, and takes approximately 14 hours for Chairs, 22 hours for Tables, and 8 hours for Storage.

Fig.~\ref{fig:gen_nn} shows some examples of novel shapes synthesized by these generative models, as well as their nearest neighbor from the training set according to both program similarity and geometric similarity.
The generative models are capable of producing valid, plausible output shapes, and they do not simply memorize their training data.

We quantitatively assess the quality of the generative models' output shapes using the following metrics (additional details in supplemental Section C):

\begin{itemize}
\denselist
    \item \textbf{Rootedness $\Uparrow$ (\% rooted):} percentage of shapes whose leaf parts all have a path to the ground.
    \item \textbf{Stability $\Uparrow$ (\% stable):} percentage of shapes which remain upright when subjected to a small vertical drop. 
    \item \textbf{Realism $\Uparrow$ (\% fool):} percentage of test set shapes classified as ``generated'' by a PointNet~\cite{qi2017pointnet} trained to distinguish between generated shapes and training set shapes.
    \item \textbf{Frechet Distance $\Downarrow$ (FD):} distributional similarity between generated shapes and training set shapes in the feature space of a pre-trained PointNet~\cite{FrechetInceptionDistance}.
\denselist
\end{itemize}

Table \ref{tab:gen_stats} shows the results of this experiment.
Metrics related to realism/plausibility (\% fool, FD) are always best for programs that use \methodname macros as opposed to other language variants.
Complexity (\# Parts) and validity (\% rooted, \% stable) metrics also generally improve.
The simple baseline macros are considerably worse; worse, in fact, than using no macros at all.
We provide some qualitative comparisons of generated outputs from \methodname vs No Macros in Appendix \ref{sec:gen_qual}.

\begin{table}[t!]
    \centering    
    \setlength{\tabcolsep}{2pt}
    \footnotesize
    \caption{
    Comparing the quality of programs sampled from a learned generative model.
    Generative models trained on programs with \methodname macros tend to produce more visually plausible, physically valid, and complex shapes than those trained on programs expressed with other libraries.
    }
    \begin{tabular}{@{}llccccc@{}}
        \toprule
        \textbf{Category} 
        & \textbf{Method} 
        & \textbf{\% fool}$\;\Uparrow$ 
        & \textbf{FD}$\;\Downarrow$
        & \textbf{\# Parts}$\;\Uparrow$ 
        & \textbf{\% rooted}$\;\Uparrow$ 
        & \textbf{\% stable}$\;\Uparrow$
        \\
        \midrule
        \multirow{3}{*}{\emph{Chair}}
        & No Macros 
        & 21.2 
        & 17.8
        & 7.6 & 
        \textbf{93.9} 
        & \textbf{82.3} 
        \\
        & Baseline Macros 
        & 16.9 
        & 24.1 
        & 8.5 
        & 89.8 
        & 74.2 
        \\
        & \methodname 
        & \textbf{25.6} 
        & \textbf{16.7}
        & \textbf{8.6} 
        & 92.7 
        & 79.5 
        \\
        \midrule
        \multirow{3}{*}{\emph{Table}}
        & No Macros 
        & 27.7 
        & 26.0 
        & \textbf{8.0}
        & 88.8 
        & 76.1 
        \\
        & Baseline Macros 
        & 11.5 
        & 38.1 
        & 7.0 
        & 90.2 
        & 79.6 
        \\
        & \methodname 
        & \textbf{29.2} 
        & \textbf{23.2} 
        & 7.8 
        & \textbf{93.2} 
        & \textbf{84.3} 
        \\
        \midrule
        \multirow{3}{*}{\emph{Storage}}
        & No Macros 
        & 4.9 
        & 70.0 
        & 6.0 
        & 92.4 
        & 85.5 
        \\
        & Baseline Macros 
        & 5.5 
        & 78.9
        & 7.6 
        & 86.2 
        & 78.3 
        \\
        & \methodname 
        & \textbf{11.1} 
        & \textbf{38.1}
        & \textbf{7.7} 
        & \textbf{95.1} 
        & \textbf{90.5} 
        \\
        \bottomrule
    \end{tabular}
    \label{tab:gen_stats}
\end{table}


\subsection{Inferring 3D Shape Structures}
\label{sec:res_infer}

\begin{figure*}[t!]
    \centering
    \setlength{\tabcolsep}{1pt}
    \begin{tabular}{ccccccc}
        \raisebox{3.0em}{\textbf{Point Cloud}  } &
        \includegraphics[{width=.14\linewidth}]{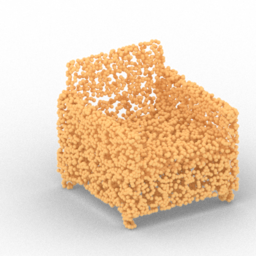} &
        \includegraphics[{width=.14\linewidth}]{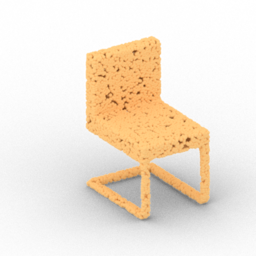} &
        \includegraphics[{width=.14\linewidth}]{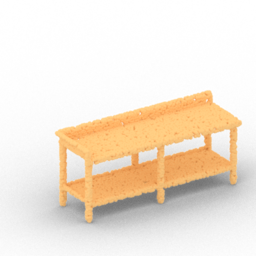} &
        \includegraphics[{width=.14\linewidth}]{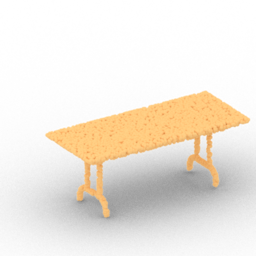} &
        \includegraphics[{width=.14\linewidth}]{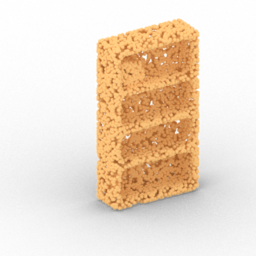} &
        \includegraphics[{width=.14\linewidth}]{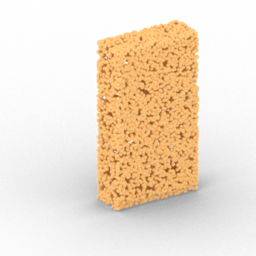} \\
        \raisebox{3.0em}{\textbf{No Macros}  } &
        \includegraphics[{width=.14\linewidth}]{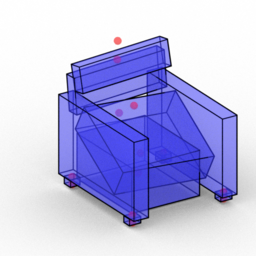} &
        \includegraphics[{width=.14\linewidth}]{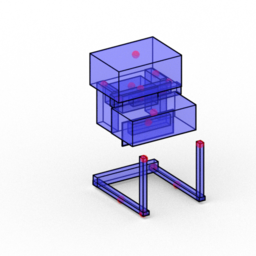} &
        \includegraphics[{width=.14\linewidth}]{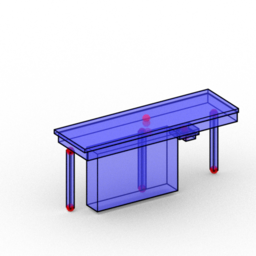} &
        \includegraphics[{width=.14\linewidth}]{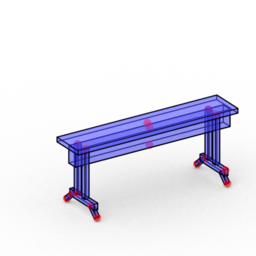} &
        \includegraphics[{width=.14\linewidth}]{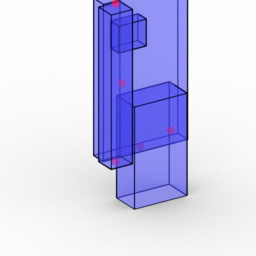} &
        \includegraphics[{width=.14\linewidth}]{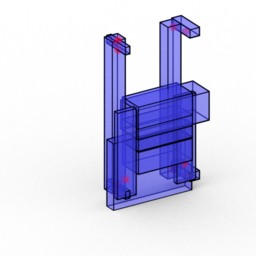} \\
        \raisebox{3.0em}{\textbf{\methodname}  } &
        \includegraphics[{width=.14\linewidth}]{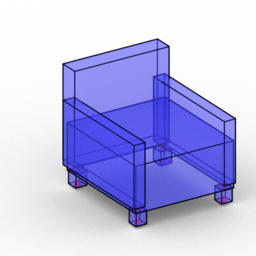} &
        \includegraphics[{width=.14\linewidth}]{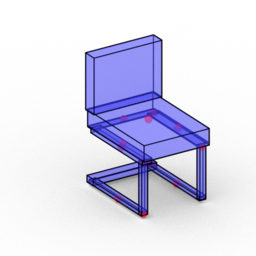} &
        \includegraphics[{width=.14\linewidth}]{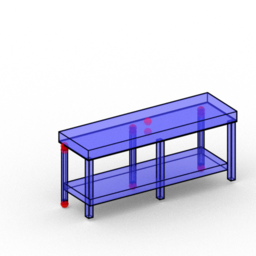} &
        \includegraphics[{width=.14\linewidth}]{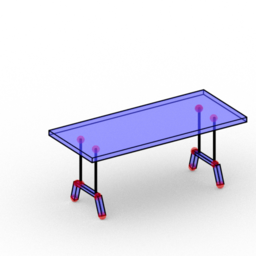} &
        \includegraphics[{width=.14\linewidth}]{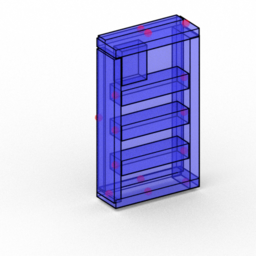} &
        \includegraphics[{width=.14\linewidth}]{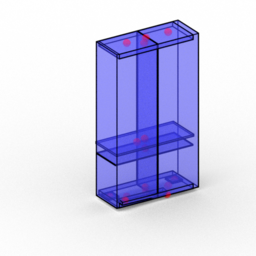} \\

    \end{tabular}
    \caption{Example visual program induction results from our point cloud $\rightarrow$ program inference experiment.
    \methodname macros are especially helpful for the heterogeneous Storage category.  All corresponding programs can be found in the supplemental material.
    }
    \label{fig:qual_prog_infer}
\end{figure*}

Another downstream task is visual program induction: inferring a shape program from unstructured input geometry.
Here, we consider inferring ShapeAssembly programs from a point cloud.
As with generative modeling, our hypothesis is that macros will regularize this problem, making it harder to output invalid shapes.

We train the program inference networks end-to-end in an encoder-decoder paradigm. 
The encoder uses a PointNet++ architecture to embed a point cloud sampled from dense surface geometry into a latent space.
The decoder is identical the one used for generative modeling, it converts a point in this latent space into a hierarchical shape program. 
We create a 80/10/10 training/validation/test set split for all categories. 
Each network is trained for 2000 epochs with a learning rate of 2e-4 and a batch size of 32. 
We report metrics on test set shapes, and choose the model that reported the best Chamfer distance on the validation set.

Table~\ref{tab:quant_prog_infer} shows the results of this experiment.
As with generative modeling, using \methodname macros results in significantly better performance.
Using \methodname macros leads to better reconstruction accuracy, in terms of Chamfer distance and F-score, for all categories (average relative improvement for both is 11\%).   
Moreover, the programs that are inferred with macros also always result in shapes that are more physically valid in terms of stability and rootedness.
Fig.~\ref{fig:qual_prog_infer} shows some example input point clouds and the shapes produced by their inferred programs.
Macros help considerably, especially for Storage, which is the most structurally- and geometrically-heterogeneous category and thus most likely to cause structured prediction models to output garbage.

\begin{table}[t!]
    \centering    
    \setlength{\tabcolsep}{2pt}
    \footnotesize
    \caption{
    Quantitative results from our visual program induction experiment, where we train encoder-decoder models that learn to infer ShapeAssembly programs from point clouds.
    \methodname macros regularize the output program space, leading to significant and consistent improvement in both reconstruction accuracy and physical validity.
    Note: Chamfer Distance (CD) values are multiplied by 1000 for clarity and we use a F-Score threshold of 0.03 \cite{TanksAndTemples}.
    }
    \begin{tabular}{@{}llcccccc@{}}
        \toprule
         \textbf{Category} &
        \textbf{Method} & \textbf{CD}$\;\Downarrow$ & \textbf{F-Score}$\;\Uparrow$  & \textbf{\% rooted}$\;\Uparrow$ & \textbf{\% stable}$\;\Uparrow$ \\
        \midrule
        \multirow{2}{*}{\emph{Chair}} 
        & No Macros & 44.2 & 54.8 & 93.7 & 83.6 \\
        &\methodname & \textbf{41.7} & \textbf{56.1} & \textbf{96.9} & \textbf{88.0} \\
        \midrule
         \multirow{2}{*}{\emph{Table}} 
        & No Macros & 41.1 & 64.0 & 92.8 & 78.2 \\
        &\methodname & \textbf{36.7} & \textbf{68.7} & \textbf{95.2} & \textbf{88.5}\\
        \midrule
         \multirow{2}{*}{\emph{Storage}} 
        & No Macros & 56.5 & 41.1 & 95.0 & 87.7\\
        &\methodname & \textbf{47.0} & \textbf{53.0} & \textbf{97.6} &\textbf{ 92.6}\\
        \bottomrule
    \end{tabular}
    \label{tab:quant_prog_infer}
\end{table}


\subsection{Interactive Shape Editing}
\label{sec:res_edit}

Our final downstream task is interactive shape editing.
We hypothesize that programs with macros will support easier, more efficient shape editing.
To test this hypothesis, we built an interactive ShapeAssembly editor and conducted a user study with it.

\begin{figure}[t!]
  \includegraphics[width=\linewidth]{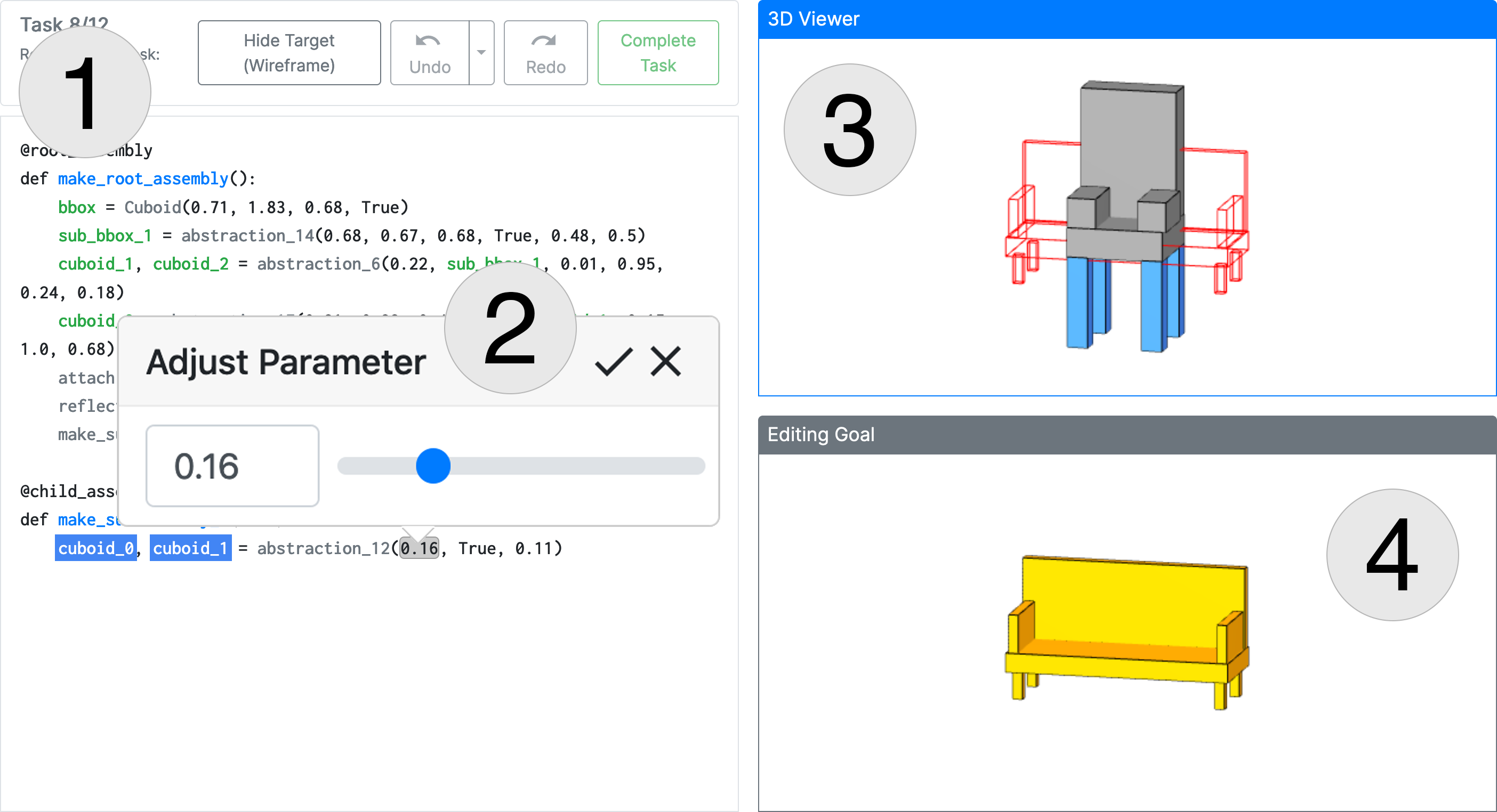}
\caption{
A screenshot of our editing interface.
The key elements are:
(1) A view of the ShapeAssembly program's text.
(2) Contextual sliders (enlarged in the figure) that allow the user to edit program parameters.
(3) A view of the current program's output. Note the optional wireframe of the target shape and the ability to highlight correspondences between cuboids in the text and the 3D viewer (blue highlights shown).
(4) The target shape.
}
\label{fig:editing_task_ui}
\end{figure}

\paragraph{Editing interface}
We designed an interactive editing interface tailored to the goal-directed editing task of modifying a ShapeAssembly program such that its output shape matches a target output shape as closely as possible.
Fig.~\ref{fig:editing_task_ui} shows our interactive editing interface.
The left panel shows the text of the current ShapeAssembly program.
The top-right panel shows the current output shape produced by this program; the bottom-right panel shows the target shape.
The cameras of the two shape view panels are synchronized, such that if a user moves the viewpoint of one, the other one follows.
The user also has the option of toggling a wireframe display of the target shape overlaid on the current output shape, which can assist with making fine-tuning edits.
Finally, in this interface, the text of the program is frozen: users are only allowed to manipulate the continuous programs parameters via contextual slider widgets that appear when a parameter is clicked.
See the supplemental video for a demonstration of the interface.

\paragraph{Experiment design}
Our study asked participants to perform a series of goal-directed editing tasks.
To ensure that it was possible to complete these tasks, we selected each target shape by finding a program in our dataset that was identical to the input program up to continuous parameters. 
We recruited 38 participants, all of whom were university students with some programming background.
Participants were randomly divided into one of two conditions: editing programs with \methodname macros or programs without them. 
\rev{Participants were not told the meaning of their assigned condition.} 
First, each participant was shown a short tutorial which explained the features of ShapeAssembly and allowed them to become familiar with the editing interface.
Then, participants completed six editing tasks (two for each of Chair, Table, and Storage).
Participants were given 10 minutes to complete each task.
After completing these tasks, participants completed an exit survey which asked them to rate the ease of each task (1-5, with 5 being easiest) as well as to provide qualitative feedback about their experience.

\paragraph{Results}

\begin{figure*}
    \centering
    \includegraphics[width=\linewidth]{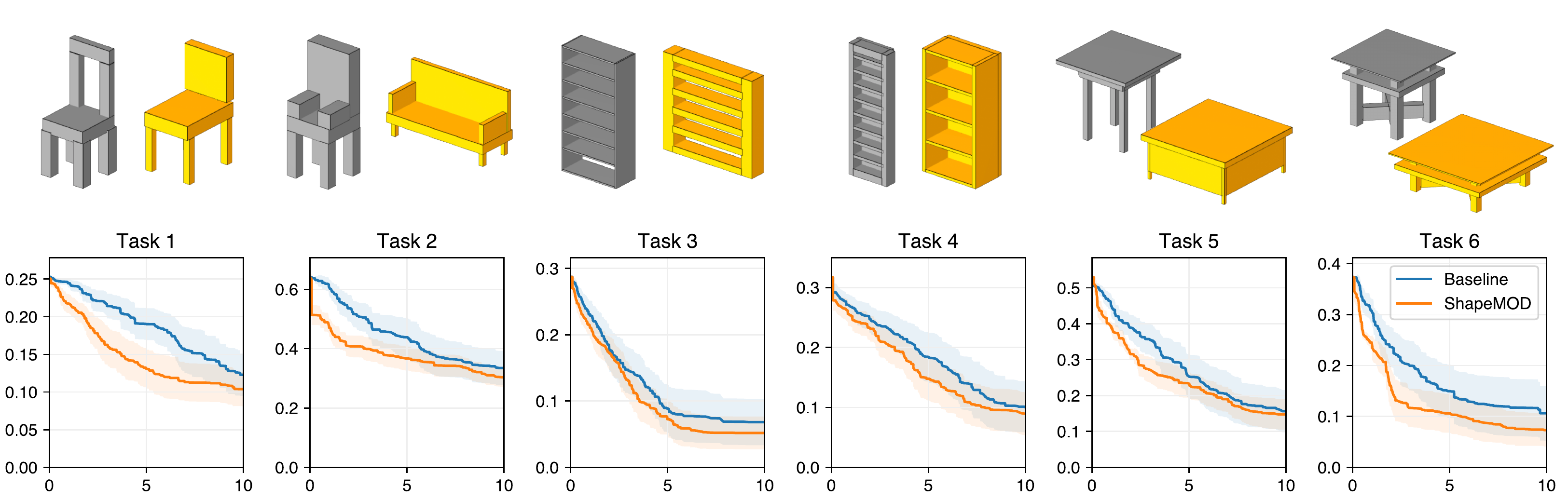}
    \caption{
    Top row: the initial program output shape (gray) and target shape (yellow) for each task in our goal-directed editing study.
    Bottom row: plots of how quickly participants were able to edit a program's parameters to match the target shape, with 95\% confidence intervals shown. The x axis is time elapsed in minutes, while the y axis is the mean of the running minimum of each participant's corner distance to the target shape.
    In general, participants using \methodname macros more quickly converged to the target shape and achieved a closer fit. To allow users to take breaks between tasks, time starts when the user makes their first edit for each task .
    }
    \label{fig:study_curves}
\end{figure*}

We first ask the question: how long did it take participants to edit the program to produce a close match to the target shape?
Fig.~\ref{fig:study_curves} plots the running lowest corner distance of the program output to the target shape as a function of task time elapsed, for each of the six study tasks, averaged across participants in each condition.
For all tasks, participants using \methodname macros more quickly converged to the target shape.

\begin{figure}
    \centering
    \includegraphics[width=0.8\linewidth]{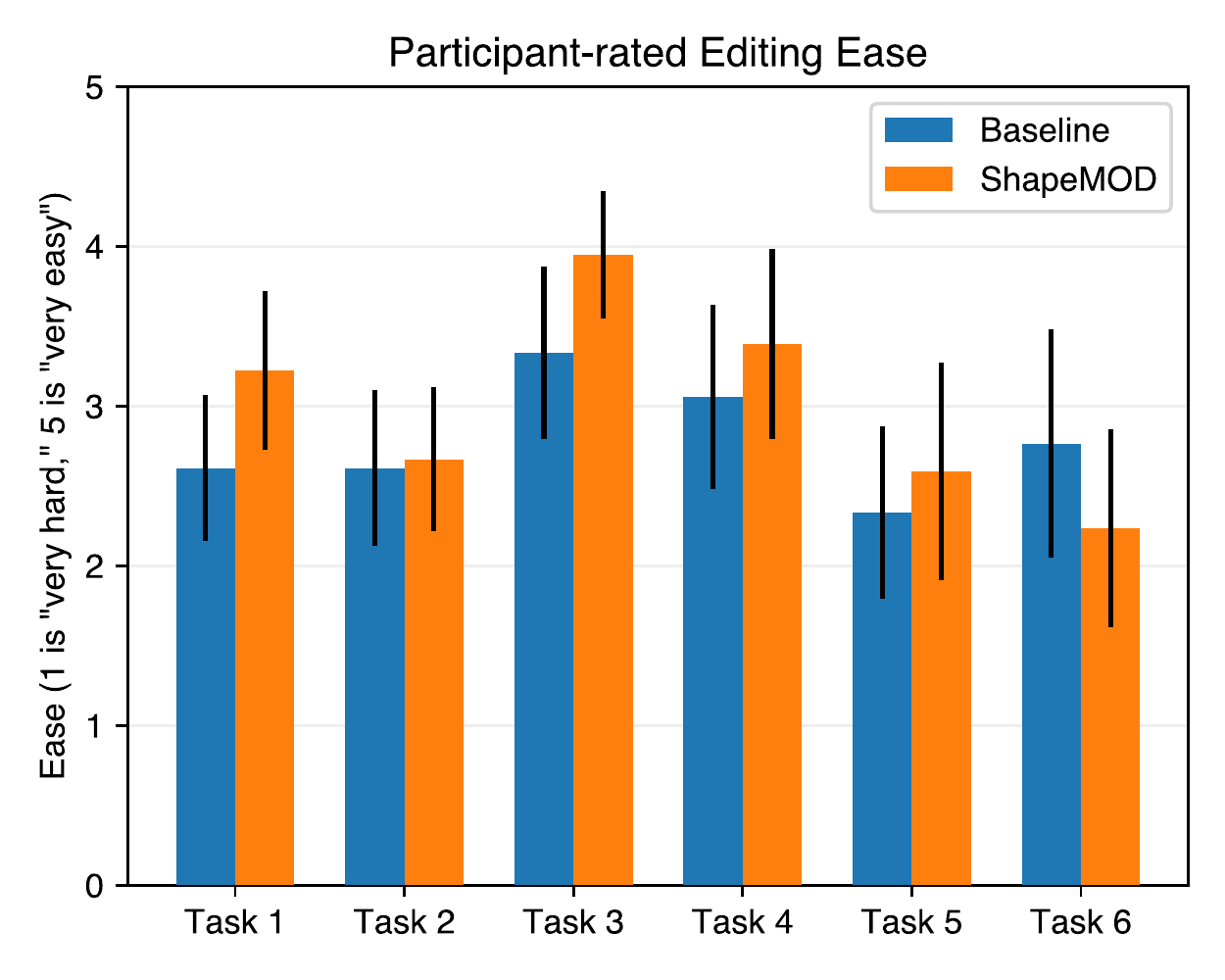}
    \caption{
    Participants in our user study rated the ease of completing each task; here, we plot each task's average difficult rating for each condition (5 = very easy, 1 = very difficult) with 95\% confidence intervals shown.
    Participants using \methodname macros generally rated tasks as easier to complete.
    }
    \label{fig:study_responses}
\end{figure}

We also examined the participants' responses to survey questions.
Fig.~\ref{fig:study_responses} shows the ease rating given to each task, averaged across participants in each condition.
For most tasks, participants using \methodname macros rated the task as slightly easier to complete.

\subsection{Cross-category Macro Discovery}

We also wondered: can one discover useful macros from a dataset consisting of multiple categories of shapes?
To answer this question, we ran the \methodname algorithm on the union of our Chair, Table, and Storage datasets, and report full quantitative results in the supplemental material (Section E). Interestingly, the library of functions discovered across multiple categories led to better program compression statistics, but slightly degraded performance on novel shape generation and program inference tasks, compared with libraries discovered by category specific \methodname runs. \rev{These experiments show that for downstream tasks it is slightly better to run \methodname on a per-category basis, although the marginal performance gap provides evidence that the discovered macros can generalize.}
\section{Conclusion \& Future Work}
\label{sec:conclusion}

We presented \methodname, an algorithm for discovering useful macros across a dataset of shape programs.
To our knowledge, \methodname is the first method that discovers common abstractions from a set of imperative programs with relationships between continuous variables.
The macros \methodname finds significantly compress the input programs, and these compressed programs lead to better results when used to train models for generating shape structures and inferring shape programs from point clouds.
We also conducted a user study which showed that compressed programs allow for more efficient shape program editing.

The abstractions that \methodname currently considers when proposing macros are relatively simple refactorings of free parameters (e.g. into constants or expressions of other variables).
The algorithm could be extended to consider other forms of abstraction, for example explaining repeated statements with for loops, use of conditional branches, etc.
Such abstractions might allow \methodname to discover even more complex macros, such as new forms of symmetry groups.

As mentioned in Section~\ref{sec:integration}, \methodname's integration step is intractable to solve optimally.
But even the greedy approximation we use can be slow for large collections of shape programs.
The major computational bottleneck is the cost of finding optimal programs $\program^*(\dataset, \library)$.
This step could potentially be accelerated via \emph{neurally-guided program search}, i.e. training a neural net which proposes which macros are most promising for explaining which lines of the original programs~\cite{lu2019neurally,ellis2018library,ellis2020dreamcoder}.

\methodname requires programs in a base DSL as input.
While some shape datasets exist which are readily convertible into such a format (e.g. PartNet~\cite{PartNet}), they are not widely available.
It would be interesting to try discovering a shape programming language (complete with macros) from scratch given, for instance, only a set of parts per shape in a shape collection.

While \methodname finds macros that are useful across shape program collections, it does not give them semantic names.
In fact, some users in our editing study found the base ShapeAssembly functions easier to work with than the macros for this reason (even though they edited more efficiently with the macros).
Finding efficient ways to assign semantic names to automatically-discovered macros would be a fruitful direction for future work.

Finally, while we  demonstrated the \methodname algorithm on ShapeAssembly programs in this paper, the method is quite general in principle: it makes no assumptions about its input language, other than that it is imperative and has certain parameter types.
Aside from shape modeling, other graphics problems may be expressible in such a language (e.g. shader programming).
We are excited to see how \methodname's automatic macro discovery capabilities might find applications elsewhere in computer graphics and beyond.
\begin{acks}
We would like to thank the participants in our user study for their contribution to our research.
We would also like to thank the anonymous reviewers for their helpful suggestions. 
Renderings of part cuboids and point clouds were produced using the Blender Cycles renderer. 
This work was funded in parts by NSF award \#1941808, a Brown University Presidential Fellowship, an ERC grant (SmartGeometry), and gifts from Adobe. 
Daniel Ritchie is an advisor to Geopipe and owns equity in the company. Geopipe is a start-up that is developing 3D technology to build immersive virtual copies of the real world with applications in various fields, including games and architecture.
\end{acks}

\bibliographystyle{ACM-Reference-Format}
\bibliography{main}

\appendix

\noindent
\newline

\section{Modified ShapeAssembly Grammar}
\label{sec:mod_sa_grammar}

\begin{table}[t]
\caption{
Modified grammar of ShapeAssembly \cite{jones2020shapeAssembly}.
}

\footnotesize
\begin{tabular}{|l|}
\hline
Start  $\xrightarrow{}$  BBoxBlock; ShapeBlock; \\
BBoxBlock $\xrightarrow{}$ $\text{bbox} = \texttt{Cuboid}(w, h, d, \texttt{True})$  \\
ShapeBlock $\xrightarrow{}$ PBlock ; ShapeBlock | None  \\
PBlock $\xrightarrow{}$  $c_n = \texttt{Cuboid}(w, h, d, a)$ ; ABlock; SBlock \\
ABlock $\xrightarrow{}$ Attach | Attach ; Attach | Squeeze \\
SBlock $\xrightarrow{}$ Reflect | Translate | None \\
Attach $\xrightarrow{}$ $\texttt{attach}(c_{n_1}, x_1, y_1, z_1, x_2, y_2, z_2)$ \\
Squeeze $\xrightarrow{}$ $\texttt{squeeze}(c_{n_1}, c_{n_2}, f, u, v)$ \\
Reflect $\xrightarrow{}$ $\texttt{reflect}(\text{axis})$ \\
Translate $\xrightarrow{}$ $\texttt{translate}(\text{axis}, m, di)$ \\

$f \xrightarrow{}$ right\: |\: left \:|\: top\: |\: bot\: |\: front\: |\: back \\
$\text{axis} \xrightarrow{}$ X\: |\: Y \:|\: Z\: \\
$w, h, d \in \reals^+$ \\
$x, y, z, u, v, di \in [0,1]^2$ \\
$a \in [\texttt{True}, \texttt{False}]$ \\
$n, m \in \integersNonNeg$ \\
\hline
\end{tabular}

\label{tab:dsl}
\end{table}

Table \ref{tab:dsl} shows the modified grammar for ShapeAssembly that we use. We make the following changes from the ShapeAssembly version presented in \cite{jones2020shapeAssembly}. Instead of having separate blocks where all cuboids are defined, then all attaches are defined, and then finally all symmetry operators are defined, we interleave the attach / symmetry commands with the cuboids they move. Specifically a program starts with defining a bounding volume, and then is followed with a series of PBlocks. Each PBlock defines a Cuboid, attaches it to at least one previous cuboid (or the bounding volume), and optionally applies a symmetry operation to it. We find that this ordering permits the discovery of more interesting and useful macros, as otherwise macros would mostly be made up of only Cuboid definitions or only attachments (instead of a mix of operators). As a by-product of this new ordering, we assume that all non-Cuboid operators (\texttt{attach}, \texttt{squeeze}, \texttt{reflect}, \texttt{translate}) always operate on the last defined cuboid, and so in this way we remove one cuboid index parameter from each of these functions.

\section{Baseline Method for Macro Operator Discovery}
\label{sec:baseline_method}

Designing a baseline for \methodname is non-trivial, because there do not exist any existing methods that are able to find macro operators over datasets of programs written in imperative languages that contain continuous parameters.  
Thus, we present a naive single-pass algorithm that mimics a simplified version of \methodname's core logic. 
It starts by choosing one order for each program in the dataset. Specifically, the most canonical order, as defined in the supplemental material (Section A.3). 
Then it records all subsequences of functions that appear in the resulting program lines. 
If any subsequence is observed in more than 10\% of programs in the dataset, then it is turned into a macro function. 
Parameters of this macro function can be converted from free parameters to constants if at least 90 \% of the parameterizations of this subsequence across the dataset had the same value (for discrete parameters) or were within .05 range of the mean value (for continuous parameters). 
Once these macros have been discovered, we use the best program finding step from \methodname to create a dataset of programs expressed with macros discovered by the baseline method. 
As shown throughout the results section, the macros discovered by \methodname outperform the macros discovered by this baseline method, for every task we consider. 

\section{A Network Architecture for any library}
\label{sec:modeling}

After running our procedure to generate a library $L$, we want to design a neural network that is able to generate programs using the functions of $L$. 
As our procedure is able to produce many different libraries $L$, depending on which macro operators it discovers, our network architecture must be flexible enough to model any set of discovered functions. 
To demonstrate that this is achievable, we generalize the neural network from \cite{jones2020shapeAssembly} so that it is able to learn how to generate programs expressed in any $L$ discovered through our procedure, and validate this works in later experiments.

The base model is a hierarchical sequence VAE. The encoder branch ingests a hierarchical program and embeds it into a high dimensional latent space. The decoder branch converts a code from this latent space into a hierarchical program. Originally, the underlying library was fixed to ShapeAssembly, so the network architecture and input representation could be tailored to one set of functions. 

We design a generalized version of this network architecture that is customized based on the library of functions $L$ discovered by our procedure.
The parts of the architecture that had to be generalized were the tensor line representation and the sub-networks in the line decoder module. 

In our new line representation, the dimension of the line tensor and meaning of each index changes depending on $L$. 
The first $|L|+2$ indices of the tensor correspond to a one hot vector denoting the function type of each line (notice we add special START and STOP tokens). 
 Then for each type of discrete parameter, $p_d$, we find its number of valid values, $p_{d\_size}$, and maximum number of $p_d$ free parameters in any function of $L$, $p_{d\_free}$. 
We then reserve $p_{d\_free}$ slots of size $p_{d\_size}$ in our tensor for $p_d$, where each slot corresponds to a one hot vector whenever $p_d$ is required by a function.
Finally, for any function $f \in L$ that takes in a set of continuous parameters, $f_c$, we reserve a slot in our tensor of size $|f_c|$.

The number and structure of sub-networks in our new line decoder model also depends on $L$. The $M_{func}$ module is responsible for predicting the line's function, and therefore has $|L| + 2$ possible outputs (the functions of $L$ and the special START and STOP tokens. For each $f \in L$, for each of its free discrete parameters $f_{d\_i}$, we add a sub-network $M_{f\_d\_i}$ responsible for predicting the ith discrete parameter of $f$. Then, for every $f$ that has free continuous parameters, we add a sub-network $M_{f\_c}$ for predicting the continuous parameters of $f$. 

We implement each sub-network as a 3 layer MLP. The network is trained in a teacher forcing paradigm with a cross entropy loss for all discrete predictions and an l1 loss for all continuous predictions. Parameter sub-networks are invoked, and tensor slots in each line are filled, depending on the function type predicted in each output line. Otherwise we use the same hyper-parameters as in Jones et al.~\shortcite{jones2020shapeAssembly}

\section{\methodname Convergence Properties}
\label{sec:convergence}

\begin{table}[]
    \centering    
    \small
    \caption{ \rev{Evaluating the convergence properties of \methodname under different ablation variants. Lower values of~\objective~are better.}}
    \begin{tabular}{@{}lc@{}}
        \toprule
        \textbf{Method} 
        & \textbf{\objective} $\Downarrow$
        \\
        \midrule
        \methodname (5 rounds) &\textbf{ 68.1} \\
        \methodname (3 rounds) & 72.0 \\
        \methodname (1 round) & 82.7 \\
        No $\programs_\textbf{matches}$ Dist (5 rounds) &  68.4\\
        No Beam Search (5 rounds) & 71.6 \\
        No Bad Order Filter (5 rounds) & 73.6 \\
        No Generalize Macros (5 rounds) & 79.3 \\
        No Valid Macros Criteria (5 rounds) & 91.5 \\
        \bottomrule
    \end{tabular}
    \label{tab:abl}
\end{table}
\rev{To characterize how different design decisions affect the convergence properties of \methodname, we ran an ablation experiment (Table \ref{tab:abl}). 
Each variant was tasked with discovering a library of macros for the dataset of chair programs, and we track how well each variant's library was able to compress the dataset according to \methodname's objective function~\objective~(Section \ref{sec:objective}).
The "$\programs_\textbf{matches}$ Dist" variant replaces the sampleByParamSim function (Alg \ref{algo:mod}, line 6) with a uniform sample over matching programs. 
The "No Beam Search" variant replaces the beam search step from Section \ref{sec:best_prog} with a best-first search approach. 
The "No Bad Order Filter" variant removes the filterBadOrders function (Alg \ref{algo:mod}, line 27). 
The "No Generalize Macros" variant removes the generalize function (Alg \ref{algo:mod}, line 9). The "No Valid Macro Criteria" variant accepts any proposed macro function, ignoring the validity criteria from Section B.2 of the supplemental. 
Notice that the full version of \methodname achieves the lowest objective function value compared with all of the other variants, and that the discovered library of macros does improve through multiple rounds of the algorithm (top three rows).}

\section{Shape Generation Qualitative Comparison}
\label{sec:gen_qual}

\begin{figure*}[t!]
    \centering
    \setlength{\tabcolsep}{1pt}
    \begin{tabular}{ccccc@{\hspace{2pt}}cccccc}
        & \multicolumn{4}{c}{\methodname} &&& \multicolumn{4}{c}{No Macros} 
        \\
        \cmidrule{2-5} \cmidrule{8-11}
        &       
        \includegraphics[width=.1\linewidth]{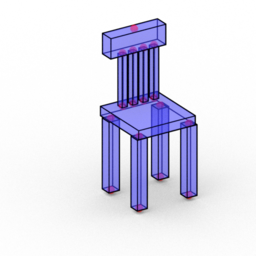} &
        \includegraphics[width=.1\linewidth]{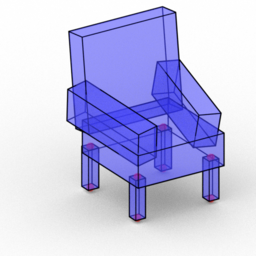} &
        \includegraphics[width=.1\linewidth]{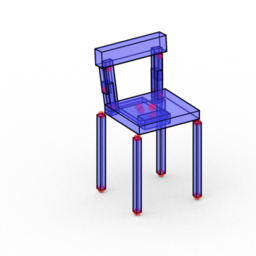} &
        \includegraphics[width=.1\linewidth]{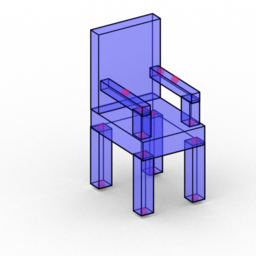}
        &&&
        \includegraphics[width=.1\linewidth]{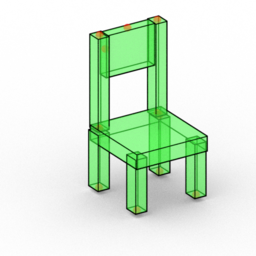} &
        \includegraphics[width=.1\linewidth]{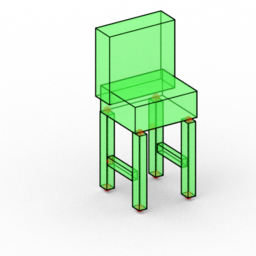} &
        \includegraphics[width=.1\linewidth]{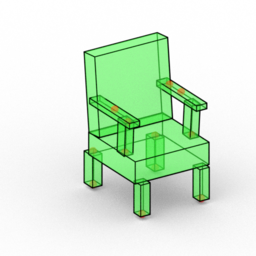} &
        \includegraphics[width=.1\linewidth]{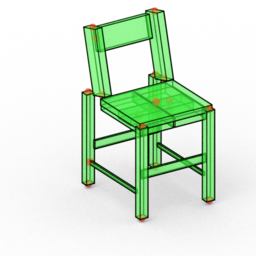}
        \\
        \raisebox{1.5em}{\rotatebox{90}{Chair}} &
        \includegraphics[width=.1\linewidth]{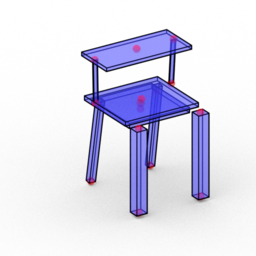} &
        \includegraphics[width=.1\linewidth]{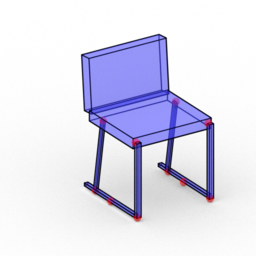} &
        \includegraphics[width=.1\linewidth]{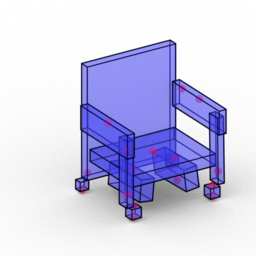} &
        \includegraphics[width=.1\linewidth]{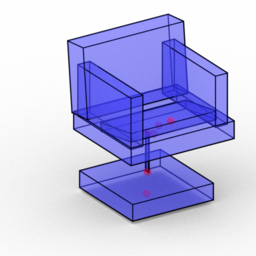}
        &&&
        \includegraphics[width=.1\linewidth]{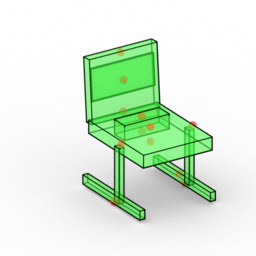} &
        \includegraphics[width=.1\linewidth]{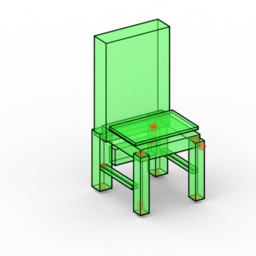} &
        \includegraphics[trim={9cm 9cm 9cm 9cm},clip,width=.1\linewidth]{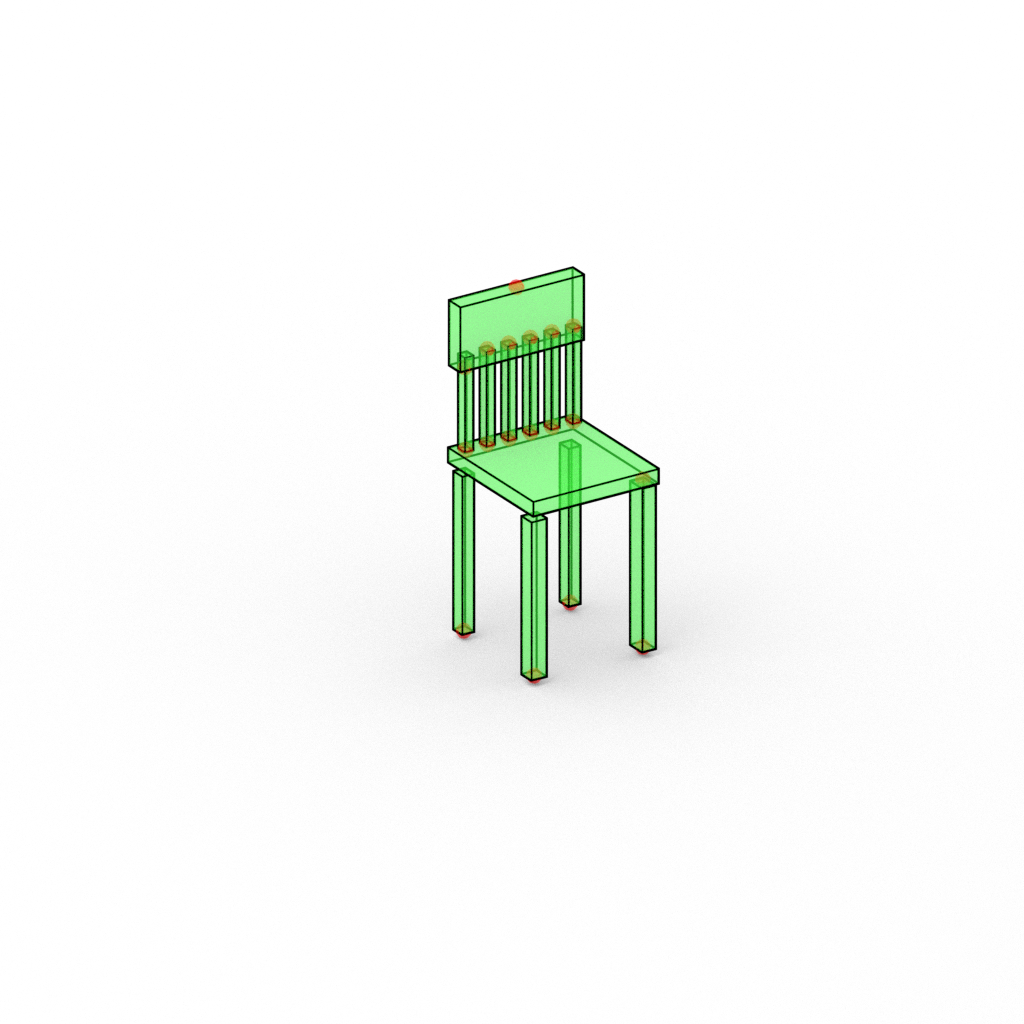} &
        \includegraphics[width=.1\linewidth]{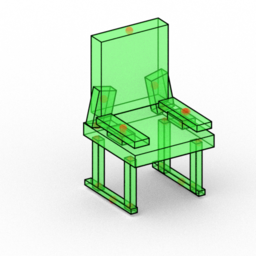}
        \\
        &
        \includegraphics[width=.1\linewidth]{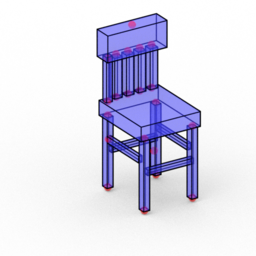} &
        \includegraphics[width=.1\linewidth]{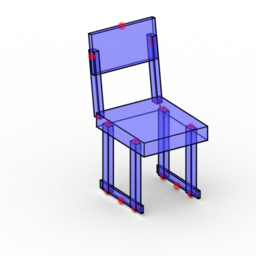} &
        \includegraphics[width=.1\linewidth]{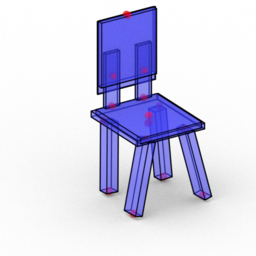} &
        \includegraphics[width=.1\linewidth]{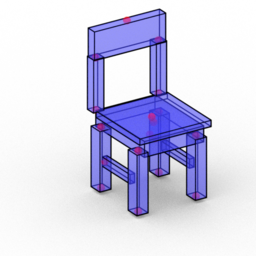}
        &&&
        \includegraphics[width=.1\linewidth]{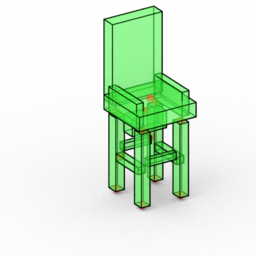} &
        \includegraphics[width=.1\linewidth]{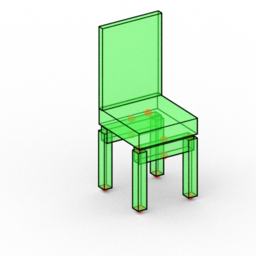} &
        \includegraphics[width=.1\linewidth]{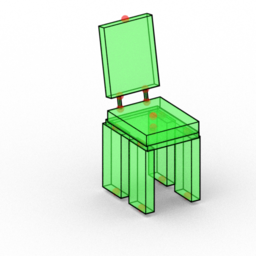} &
        \includegraphics[width=.1\linewidth]{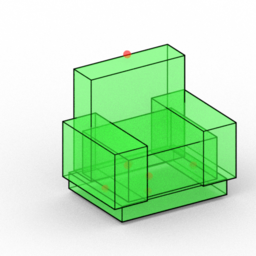}
        \\
        & \multicolumn{4}{c}{\methodname} &&& \multicolumn{4}{c}{No Macros} \\
        \cmidrule{2-5} \cmidrule{8-11}
        & \includegraphics[width=.1\linewidth]{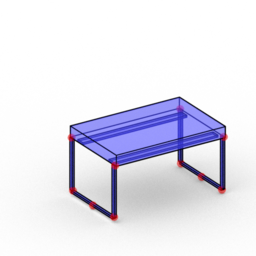} &
        \includegraphics[width=.1\linewidth]{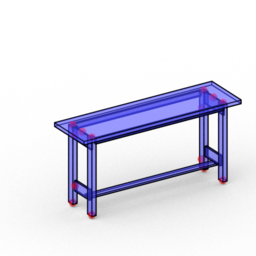} &
        \includegraphics[width=.1\linewidth]{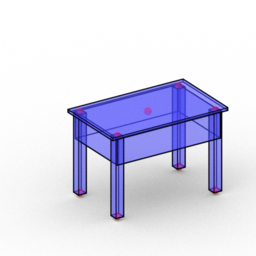} &
        \includegraphics[width=.1\linewidth]{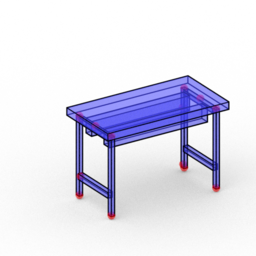}
        &&&
        \includegraphics[width=.1\linewidth]{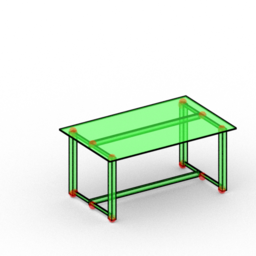} &
        \includegraphics[width=.1\linewidth]{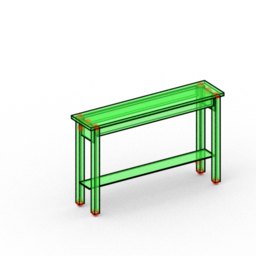} &
        \includegraphics[width=.1\linewidth]{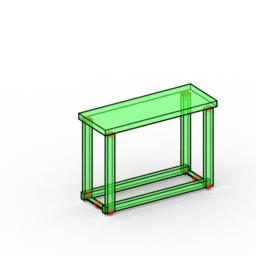} &
        \includegraphics[width=.1\linewidth]{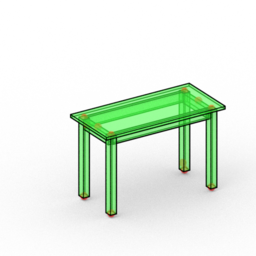}
        \\
        \raisebox{1.5em}{\rotatebox{90}{Table}} &
        \includegraphics[width=.1\linewidth]{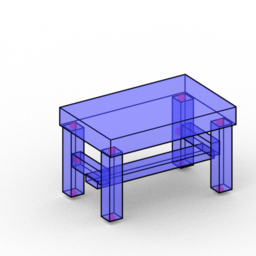} &
        \includegraphics[width=.1\linewidth]{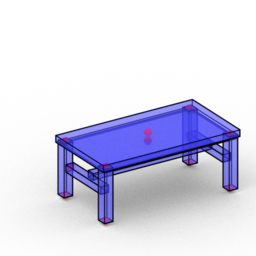} &
        \includegraphics[width=.1\linewidth]{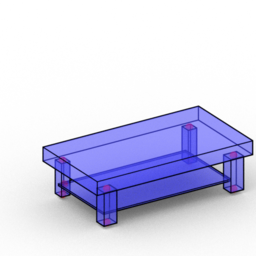} &
        \includegraphics[width=.1\linewidth]{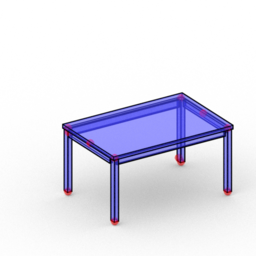}
        &&&
        \includegraphics[width=.1\linewidth]{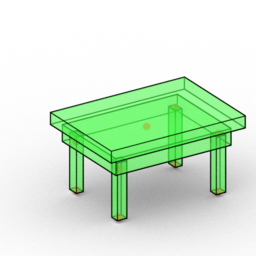} &
        \includegraphics[width=.1\linewidth]{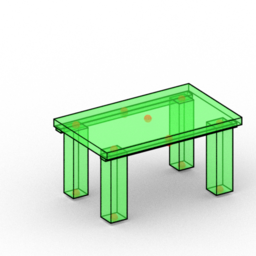} &
        \includegraphics[width=.1\linewidth]{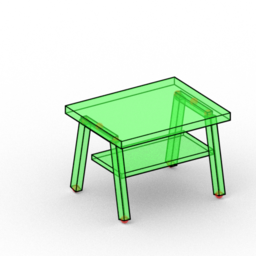} &
        \includegraphics[width=.1\linewidth]{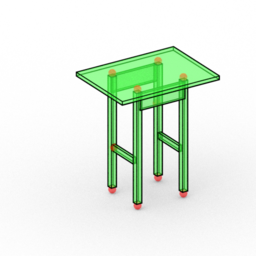}
        \\
        &
        \includegraphics[width=.1\linewidth]{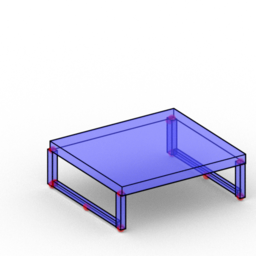} &
        \includegraphics[width=.1\linewidth]{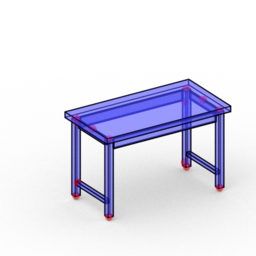} &
        \includegraphics[width=.1\linewidth]{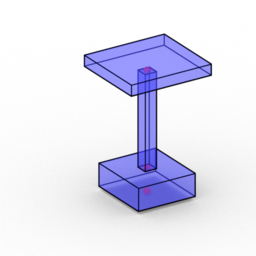} &
        \includegraphics[width=.1\linewidth]{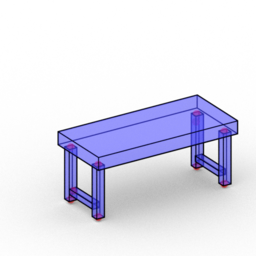}
        &&&
        \includegraphics[width=.1\linewidth]{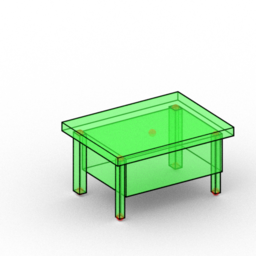} &
        \includegraphics[width=.1\linewidth]{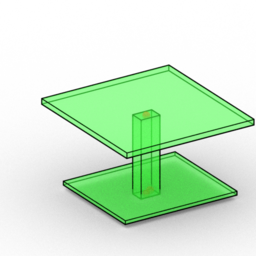} &
        \includegraphics[width=.1\linewidth]{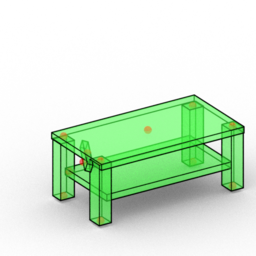} &
        \includegraphics[width=.1\linewidth]{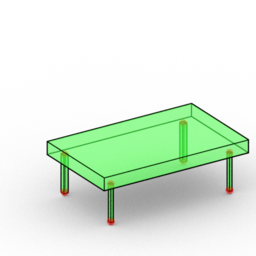}
        \\
        & \multicolumn{4}{c}{\methodname} &&& \multicolumn{4}{c}{No Macros} \\
        \cmidrule{2-5} \cmidrule{8-11}
        & \includegraphics[width=.1\linewidth]{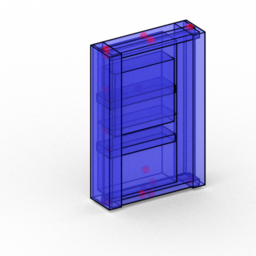} &
        \includegraphics[width=.1\linewidth]{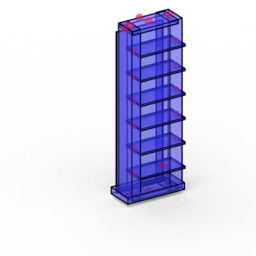} &
        \includegraphics[width=.1\linewidth]{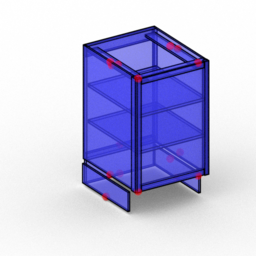} &
        \includegraphics[width=.1\linewidth]{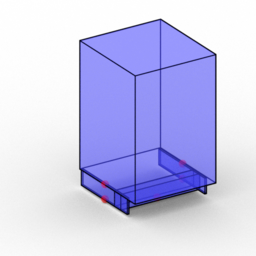}
        &&&
        \includegraphics[width=.1\linewidth]{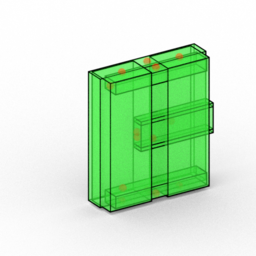} &
        \includegraphics[width=.1\linewidth]{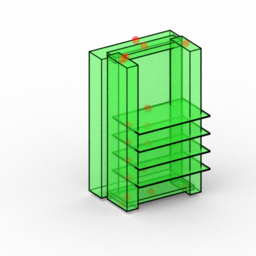} &
        \includegraphics[width=.1\linewidth]{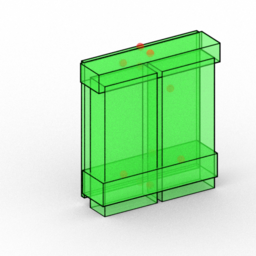} &
        \includegraphics[width=.1\linewidth]{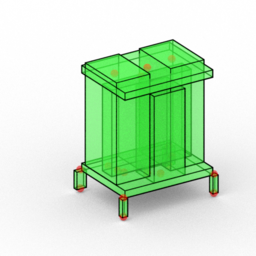}
        \\
        \raisebox{1.5em}{\rotatebox{90}{Storage}} &
        \includegraphics[width=.1\linewidth]{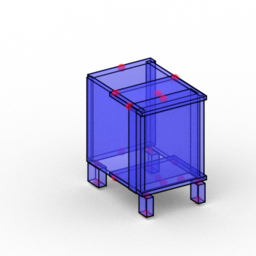} &
        \includegraphics[width=.1\linewidth]{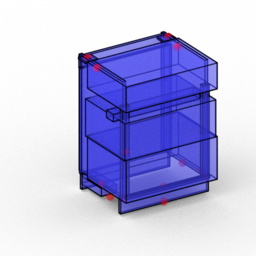} &
        \includegraphics[width=.1\linewidth]{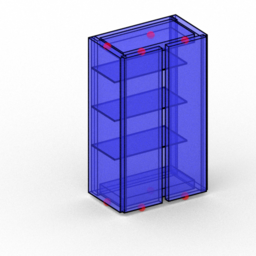} &
        \includegraphics[width=.1\linewidth]{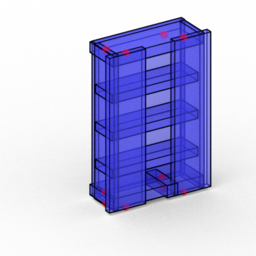}
        &&&
        \includegraphics[width=.1\linewidth]{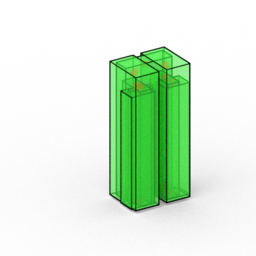} &
        \includegraphics[width=.1\linewidth]{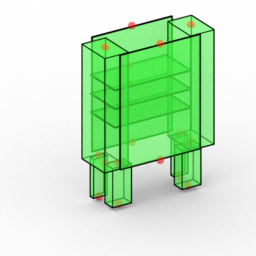} &
        \includegraphics[width=.1\linewidth]{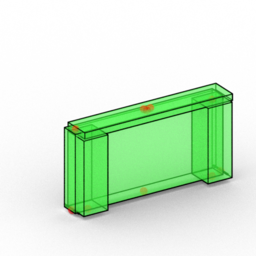} &
        \includegraphics[width=.1\linewidth]{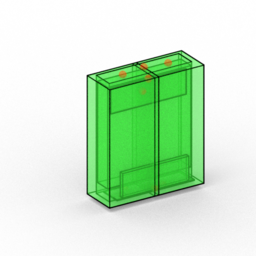}
        \\
        &
        \includegraphics[width=.1\linewidth]{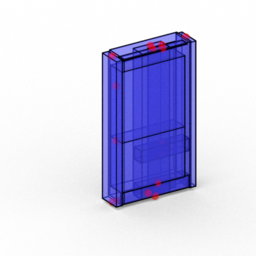} &
        \includegraphics[width=.1\linewidth]{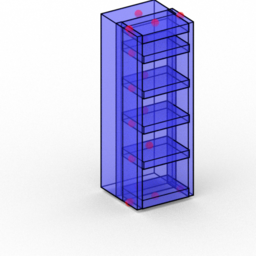} &
        \includegraphics[width=.1\linewidth]{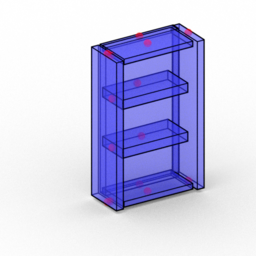} &
        \includegraphics[width=.1\linewidth]{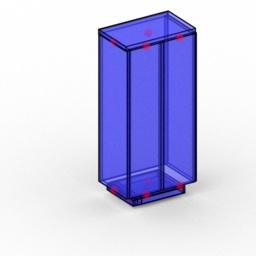}
        &&&
        \includegraphics[width=.1\linewidth]{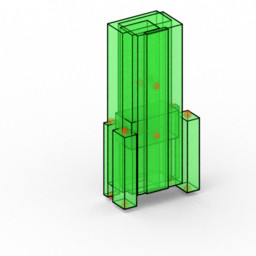} &
        \includegraphics[width=.1\linewidth]{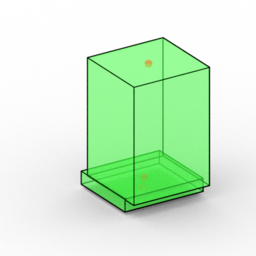} &
        \includegraphics[width=.1\linewidth]{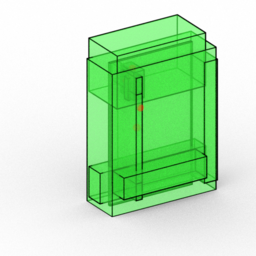} &
        \includegraphics[width=.1\linewidth]{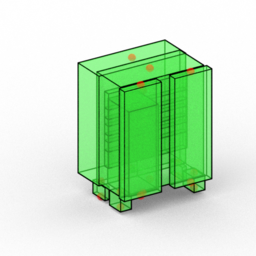}
        
    \end{tabular}
    \caption{Samples generated from generative models of ShapeAssembly programs with \methodname macros (blue) and without macros (green).
    }
    \label{fig:qual_comparison}
\end{figure*}

We share some interesting representative shape programs output by learned ShapeAssembly generative models in Figure \ref{fig:qual_comparison}. Outputs by the model trained with \methodname macros are shown in blue. Outputs by the model trained on the No Macros programs are shown in green. We include all generated program text in the supplemental materials. These qualitative results enforce the trends of our earlier quantitative experiments from Section \ref{sec:res_gen}. The best generations from Chairs and Tables are qualitatively similar, although across entire shape collections we calculated that programs with \methodname macros were more plausible. 
For storage, the qualitative difference is more pronounced, as the generations that use \methodname macros are able to create output shapes that are much closer in distribution to the target shape collections. 

\end{document}